 \documentclass{emulateapj}
 \journalinfo{Accepted for publication in the ApJ on Sep 16, 2013}
\usepackage{multirow,natbib,subfigure} 

\newcommand{\myemail}{inami@noao.edu}

\slugcomment{Not to appear in Nonlearned J., 45.}

\shorttitle{MIR Spectra of Luminous Infrared Galaxies}
\shortauthors{Inami et al.}

\begin{document}


\title{Mid-Infrared Atomic Fine-Structure Emission Line Spectra of
  Luminous Infrared Galaxies: Spitzer/IRS Spectra of the GOALS Sample}


\author{
  H. Inami \altaffilmark{1,2},  
  L. Armus \altaffilmark{2}, 
  V. Charmandaris \altaffilmark{3,4,5}, 
  B. Groves \altaffilmark{6}, 
  L. Kewley \altaffilmark{7}, 
  A. Petric \altaffilmark{8}, 
  S. Stierwalt \altaffilmark{9,2}, 
  T. D\'iaz-Santos \altaffilmark{2}, 
  J. Surace \altaffilmark{2}, 
  J. Rich \altaffilmark{10}, 
  S. Haan \altaffilmark{11}, 
  J. Howell \altaffilmark{2}, 
  A. Evans \altaffilmark{9,12}, 
  J. Mazzarella \altaffilmark{13}, 
  J. Marshall \altaffilmark{2}, 
  P. Appleton \altaffilmark{14}, 
  S. Lord \altaffilmark{13}, 
  H. Spoon \altaffilmark{15}, 
  D. Frayer \altaffilmark{16}, 
  H. Matsuhara \altaffilmark{17}, 
  S. Veilleux \altaffilmark{18} 
}
\email{\myemail}

\altaffiltext{1}{National Optical Astronomy Observatory, Tucson, AZ 85719, USA}
\altaffiltext{2}{Spitzer Science Center, California Institute of Technology, CA 91125, USA}
\altaffiltext{3}{Department of Physics and Institute of Theoretical \& Computational Physics, University of Crete, GR-71003, Heraklion, Greece}
\altaffiltext{4}{IESL/Foundation for Research \& Technology-Hellas, GR-71110, Heraklion, Greece}
\altaffiltext{5}{Chercheur Associ\'{e}, Observatoire de Paris, F-75014,  Paris, France}
\altaffiltext{6}{Max Planck Institute for Astronomy, K\"{o}nigstuhl 17,
  D-69117 Heidelberg, Germany}
\altaffiltext{7}{Research School of Astronomy and Astrophysics, The Australian National University, Cotter Road, Weston Creek, ACT 2611, Australia}
\altaffiltext{8}{Department of Astronomy, California Institute of Technology, MS 320-47, Pasadena, CA 91125, USA}
\altaffiltext{9}{Department of Astronomy, University of Virginia, P.O. Box 400325, Charlottesville, VA 22904, USA}
\altaffiltext{10}{The Observatories, Carnegie Institute of Washington, 813 Santa Barbara Street, Pasadena, CA 91101, USA}
\altaffiltext{11}{CSIRO Astronomy \& Space Science, Marsfield NSW 2122, Australia}
\altaffiltext{12}{National Radio Astronomy Observatory, 520 Edgemont Road, Charlottesville, VA 22903, USA}
\altaffiltext{13}{Infrared Processing \& Analysis Center, MS 100-22, California Institute of Technology, Pasadena, CA 91125, USA}
\altaffiltext{14}{NASA Herschel Science Center, 770 S. Wilson Ave., Pasadena, CA 91125, USA}
\altaffiltext{15}{Cornell University, Astronomy Department, Ithaca, NY 14853, USA}
\altaffiltext{16}{National Radio Astronomy Observatory, P.O. Box 2, Green Bank, WV 24944, USA}
\altaffiltext{17}{Institute of Space and Astronautical Science, Japan Aerospace Exploration Agency, Japan}
\altaffiltext{18}{Astronomy Department, University of Maryland, College Park, MD 20742, USA}


\begin{abstract}

  We present the data and our analysis of mid-infrared atomic
  fine-structure emission lines detected in Spitzer/IRS
  high-resolution spectra of 202 local Luminous Infrared Galaxies
  (LIRGS) observed as part of the Great Observatories All-sky LIRG
  Survey (GOALS).  We readily detect emission lines of [SIV], [NeII],
  [NeV], [NeIII], [SIII]$_{18.7 \, {\rm {\mu m}}}$, [OIV], [FeII],
  [SIII]$_{33.5 \, {\rm {\mu m}}}$, and [SiII]. More than $75$\% of
  these galaxies are classified as starburst dominated sources in the
  mid-infrared, based on the [NeV]/[NeII] line flux ratios and
  equivalent width of the $6.2 \mathrm{\mu m}$ polycyclic aromatic
  hydrocarbon feature.  We compare ratios of the emission line fluxes
  to those predicted from stellar photo-ionization and
  shock-ionization models to constrain the physical and chemical
  properties of the gas in the starburst LIRG nuclei.
  Comparing the [SIV]/[NeII] and [NeIII]/[NeII] line ratios to the
  Starburst99-Mappings~III models with an instantaneous burst history,
  the emission line ratios suggest that the nuclear starbursts in our
  LIRGs have ages of $1-4.5$~Myr, metallicities of $1-2$~$Z_\odot$,
  and ionization parameters of $2-8 \times 10^7 \,
  \mathrm{cm\,s^{-1}}$.  Based on the [SIII]$_{33.5 \, {\rm {\mu
        m}}}$/[SIII]$_{18.7 \, {\rm {\mu m}}}$ ratios, the electron
  density in LIRG nuclei is typically one to a few hundred cm$^{-3}$,
  with a median electron density of $\sim 300 \, \mathrm{cm^{-3}}$,
  for those sources above the low density limit for these lines.
  We also find that strong shocks are likely
  present in 10 starburst dominated sources of our sample.  
  A significant fraction of the GOALS sources (80) have resolved neon
  emission line profiles (FWHM $\ge 600\,{\rm km\,s^{-1}}$) and five
  show clear differences in the velocities of the [NeIII] or [NeV]
  emission lines, relative to [NeII], of more than $200\,{\rm
    km\,s^{-1}}$.  Furthermore, six starburst and five AGN dominated
  LIRGs show a clear trend of increasing line width with ionization
  potential, suggesting the possibility of a compact energy source and
  stratified ISM in their nuclei.
  We confirm a strong correlation between the sum of the
  [NeII]$_{12.8\,{\rm \mu m}}$ and [NeIII]$_{15.5\,{\rm \mu m}}$
  emission, as well as [SIII]$_{33.5\,{\rm \mu m}}$, with both the
  infrared luminosity and the $24\,\mu$m warm dust emission measured
  from the spectra, consistent with all three lines tracing ongoing
  star formation.
  Finally, we find no correlation between the hardness of the
  radiation field or the emission line width and the ratio of the
  total infrared to $8\,{\rm \mu m}$ emission (IR8), a measure of the
  strength of the starburst and the distance of the LIRGs from the
  star-forming main-sequence.  This may be a function of the fact that
  the infrared luminosity and the mid-infrared fine-structure lines are sensitive to different timescales
  over the starburst, or that IR8 is more sensitive to the geometry of 
  the region emitting the warm dust than the radiation field producing 
  the HII region emission.
\end{abstract}





\section{Introduction}\label{sec:intro}


Luminous infrared galaxies (LIRGs) and ultra luminous infrared
galaxies (ULIRGs) are defined to have $8-1000\,\mu$m infrared
luminosities ($L_{IR}$) of $\mathrm{log}(L_{IR}/L_\odot) \ge 11$ and
$\mathrm{log}(L_{IR}/L_\odot) \ge 12$, respectively.
The local space density of (U)LIRGs is relatively low, but they play a
central role in the understanding of the star formation history and
the evolution of black holes in the Universe.  The number density of
(U)LIRGs increases rapidly between $0 < z < 1$
\citep[e.g.,][]{Berta11}. By $z \sim 1$, LIRGs contribute at least
$50\%$ of the total comoving star formation rate density in the
Universe \citep[e.g.,][]{Seym09,Magn13}. Despite the
importance of high-z LIRGs and ULIRGs as a key source of the cosmic
infrared background radiation, it is only recently that local complete
samples have become the target of extensive, multi-wavelength studies.


The Great Observatories All-sky LIRG Survey
\citep[GOALS~\footnote{Visit the GOALS website at
  http://goals.ipac.caltech.edu};][]{Armu09} targets 202 LIRGs in the
local Universe drawn from the IRAS Revised Bright Galaxy Sample
\citep[RBGS;][]{Sand03}, which covers galactic latitudes greater than
five degrees and includes 629 extragalactic objects with
$60\,\mathrm{\mu m}$ flux densities greater than 5.24~Jy. The range of
the luminosity distances ($d_{L}$) and the median $d_L$ of the GOALS
sample are $15.9 \leq d_{L} \, \mathrm{[Mpc]} \leq 400$ and
$94.3$~Mpc, respectively.  The GOALS sample covers the full range of
galaxy interaction stages from single isolated disk galaxies, widely
separated pairs, to final stage mergers.
GOALS is an ideal sample for identifying interesting and rare sources,
and most importantly, for creating an accurate picture, through the
use of multi-wavelength probes of the gas and dust, of the
relationship between star formation and the growth of black holes
during periods of intense activity and galactic mergers. Using a
number of mid-infrared diagnostics, \cite{Petr11} estimate the active
galactic nuclei (AGN) contribution to the mid-infrared luminosity of
the systems as a function of merger stage, infrared luminosity, and
spectral shape.  A key result of this work is that the vast majority
of local LIRGs are dominated by ongoing star formation.  While there
are indeed a good number of extremely powerful AGN and composite
(starburst plus AGN) galaxies in our sample, most of the LIRGs in the
local Universe derive the bulk of their power through star formation.
In \cite{Howe10} the relation between the UV and far-infrared emission
of the sample is studied and compared to normal galaxies, while
\cite{Haan11a} present a thorough analysis of the nuclear structure of
the galaxies using high spatial resolution near-infrared and optical
images obtained with the Hubble Space Telescope. The low resolution
Spitzer mid-infrared spectra of the GOALS sample are ideal for
studying the spectral shapes and dust emission and absorption
features. They have been presented in a series of papers by
\cite{Diaz10b,Diaz11} and \cite[][and in prep.]{Stie13a}, in
addition to \cite{Petr11} as described above.  We refer the reader to
these papers since we will rely on their findings for the
interpretation of our results below.  Here, we present the results
from the fitting of the atomic, fine-structure emission lines in the
Infrared Spectrograph \citep[IRS;][]{Houc04} high resolution
short-High (SH) and Long-High (LH) for the complete GOALS sample.  We
present line fluxes, line widths and line centroids, and determine the
range of properties of the starburst-dominated LIRGs, and compare
these to photo- and shock-ionized models.

One of the most powerful ways to reveal the physical conditions in the
gas in and around the dust enshrouded nuclei of LIRGs, is to use the
dust-penetrating power of mid-infrared spectroscopy.  In particular,
the high resolution modules of the IRS on the Spitzer Space Telescope
provide access to a suite of fine structure emission lines that can be
used to diagnose the central energy source and the physical and
chemical conditions of the gas in LIRG nuclei (e.g., the hardness of
the radiation field, the gas density, and metallicity).  These
properties can then be compared to models to estimate the starburst
age or velocity of a shock.  As a result mid-infrared spectra from the
Infrared Space Observatory as well as from the Spitzer Space Telescope
have been successfully used to map the ionization field in many types
of galaxies \citep[e.g.,][]{Thor00, Stur02, Verm03, Wu06, Dale06,
  Farr07, Grov08, Bern09, Hao09, Veil09a}.

Several theoretical models are available for studying the radiation
field and the interstellar medium of galaxies
\citep[e.g.,][]{Leit99,Dopi00,Kewl01,Snij07,Leve10}.  While these
models have been used to analyze both active (AGN) and starburst
galaxies \citep{Stur02,Verm03,Bern09,Tomm10,Pere10b}, complete samples
of local LIRGs, of the size of GOALS and with the same abundance of
ancillary data, have thus far escaped detailed investigation.  In this
paper, we attempt to constrain the physical conditions in the
starburst galaxies that make up the GOALS sample, via a comparison of
the mid-infrared emission line properties to other data, and recent
models.


The paper is organized as follows. In Section~\ref{sec:obs}, the
details of Spitzer IRS observations and data reduction are
presented. In Section~\ref{sec:models}, the starburst and shock models
that we used are described. This is followed by the results in
Section~\ref{sec:results}, where the tables of our measurements for
the whole sample are presented, along with a comparison with line
ratios predicted by the models and those seen in other samples.
Finally, the findings are summarized in
Section~\ref{sec:summary}. Throughout this paper, an $H_0 = 70 \,
\mathrm{km \, s^{-1} \,Mpc^{-1}}$, $\Omega_m = 0.28$, and
$\Omega_\Lambda = 0.72$ are assumed.


\section{Observations and Data Reduction}\label{sec:obs}

The high resolution spectra of the GOALS sources were obtained with
Spitzer/IRS. In total, there are 244 and 246 nuclei observed with the
SH and LH IRS modules, respectively, within 202 LIRG systems.  For
LIRGs with multiple nuclei, secondary (fainter) nuclei were observed
with the IRS only when their flux density at $24\,{\rm \mu m}$ (as
measured from Spitzer MIPS imaging) was equal to or greater than $1/5$
the $24\,{\rm \mu m}$ flux density of the primary (brighter) nucleus.
The SH and LH spectra cover a range of $10-20\,\mathrm{\mu m}$ and
$19-38\,\mathrm{\mu m}$, respectively.  The bulk of the IRS
observations ($152/246$) were carried out under program 30323 (PI:
Armus), with the remainder of the data retrieved from the Spitzer
archive.  There is a large range in on-source integration times, from
$30 \,{\rm sec}$ to $610 \,{\rm sec}$, depending on the brightness of
the nuclei.

The data were reduced using the S15, S16, and S18 IRS pipelines at the
Spitzer Science Center.  The pipeline software removes bad pixels and
droop, subtracts the background, corrects for non-linearity, and
performs a wavelength and flux calibration.  Dedicated backgrounds
were subtracted for all data where available, in order to provide the
cleanest removal of time varying bad pixels.  One dimensional spectra
were extracted with the
SPICE~\footnote{http://irsa.ipac.caltech.edu/data/SPITZER/docs/dataanalysistools/tools/spice/}
software package which performs a full slit extraction for the SH and
LH data and applies slit loss corrections assuming a point source
dominates the flux.  This is appropriate since we are interested here
in the nuclear starbursts on scales of a few kpc, which drive the bulk
of the $L_{IR}$ emission - see below.

All line fluxes were measured by fitting a Gaussian to the IRS
emission line profiles.  Final line fluxes were derived as a
weighted-mean of the line fluxes from each spectral nod position ---
in standard IRS Staring mode observations the target is placed at two
slit positions.  When the line was not detected, $3\sigma$ upper
limits were evaluated using a Gaussian with a height three times the
local continuum dispersion and a full width at half maximum (FWHM) of
$\sim 500\,\mathrm{km\,s^{-1}}$ (the intrinsic resolution of IRS).  In
order to estimate the effective measured spectral resolution at each
wavelength (including the effects of pointing, undersampling, and
spectral extraction), we have made histograms of the measured
(Gaussian) line widths for each emission line for the entire sample.
This distribution in measured line widths shows a sharp peak for each
line, which we take as the effective resolution for that transition.
These are 500, 505, and 520 ${\rm km\,s^{-1}}$ for [NeII], [NeIII],
and [NeV] respectively.  We use the standard deviations of the
distributions in line width, which are 37, 33 and 98 ${\rm
  km\,s^{-1}}$ for [NeII], [NeIII], and [NeV] respectively, as
measures of the uncertainties in the resolution of each line, and we
add these to the fitting errors, in quadrature, for each line in a
given galaxy.


\section{Models}\label{sec:models}

A primary goal of this paper is to use the complete set of
mid-infrared fine structure lines in order to determine the physical
properties and the chemical abundances of the gas in the GOALS nuclei.
To achieve this, we compare sets of observed line ratios to those
predicted by the photoionization models of \cite{Leve10} to derive
starburst ages, electron densities, gas-phase metallicities, and
ionization parameters, as well as the shock models of \cite{Alle08},
to derive shock speeds and magnetic field strengths.  The parameter
space explored by these models is described in more detail below.

\subsection{Photoionization Model}\label{subsec:phot_model}

We have used the photoionization model of \cite{Leve10} based on the
Starburst99-Mappings~III code (versions 5.1 and 3q,
respectively~\footnote{Provided at
  http://www.emlevesque.com/model-grids/}). In this case, the
Starburst99 stellar population synthesis code \citep{Leit99, Vazq05}
is used to generate stellar spectral energy distributions (SEDs),
which are used as inputs to the Mappings~III photoionization code
\citep{Bine85, Suth93, Grov04a} to predict the resulting nebula
emission.  We briefly summarize below the parameters assumed with the
Levesque models. For more details we refer to \cite{Leve10}.

Following \cite{Leve10}, we use a Salpeter initial mass function
\citep[IMF;][]{Salp55} with an upper mass boundary of
$100\,\rm{M_\odot}$, and the stellar atmosphere models of
\cite{Paul01} and the \cite{Hill98}.  \cite{Leve10} then compute
models for two different sets of Geneva stellar evolutionary tracks
with ``Standard'' and ``High'' mass-loss rates. In this paper, we
apply only the ``Standard'' Geneva mass-loss evolutionary tracks, as
suggested by \cite{Leve10} under consideration for the effect of wind
clumping on mass-loss rates \citep{Crow02}. The final stellar
population spectra are determined for five metallicities ($Z$;
$0.05Z_\odot$, $0.02Z_\odot$, $0.4Z_\odot$, $Z_\odot$, and
$2Z_\odot$), and two star formation histories (continuous and
instantaneous burst), and sampled every $0.5$~Myr.

For each model age and metallicity, the Mappings~III models are used
to generate the final nebular emission line spectra, including the
mid-infrared lines we examine in this work. An isobaric plane-parallel
geometry for the gas is assumed, and two mean electron densities
($n_e$; $10\,\mathrm{cm^{-3}}$ and $100\,\mathrm{cm^{-3}}$), together
with seven ionization parameters ($q$~\footnote{This is defined as $U
  \equiv q/c $ where $U$ is the dimensionless ionization parameter.};
$1\times10^7$, $2\times10^7$, $4\times10^7$, $8\times10^7$,
$1\times10^8$, $2\times10^8$, and $4\times10^8 \, {\rm cm \, s^{-1}}$)
are modeled. The metallicity of the gas is matched to the stars,
assuming the same abundance pattern as detailed in \cite{Grov04a}.

This range of metallicities, star formation histories, ages,
densities, and ionization parameters should cover the expected
parameter space of the star forming regions providing the nebula
emission in the GOALS sample.  While other variables such as the IMF
and differing stellar evolutionary tracks can affect the ionizing SED,
and thus the final emission-line spectrum, exploration of these
parameters lies beyond the scope of this work.

\subsection{Shock-ionization Model}

We also used the shock models of \cite{Alle08} based on the Mappings
III shock code (version 3q), to explore the possible presence of
shock-ionized gas in the GOALS sources.  Note that the models by
\cite{Alle08} do not include any treatment of dust, but this is often
taken as a reasonable approximation given that the dust grains can be
destroyed by fast shocks.  The model parameters are pre-shock
densities, abundances, magnetic parameters, and shock velocities.  In
this investigation, we adopt a shock model with a pre-shock density
$n_e=100\,\mathrm{cm^{-3}}$ and solar metallicity, as will be
described in the next section.


\section{Results}\label{sec:results}

Our Spitzer IRS SH and LH measurements for the entire GOALS sample are
presented in Tables~\ref{tbl:spec_SH} and \ref{tbl:spec_LH},
respectively. We readily detect [SIV], [NeII], [NeV], [NeIII], and
[SIII]$_{18.7 \, {\rm {\mu m}}}$, in the $10-20 \, {\rm \mu m}$ SH
wavelength range, while between $19$ and $37 \, {\rm \mu m}$ covered
by LH, we identify [SIII]$_{18.7 \, {\rm {\mu m}}}$, [OIV], [FeII],
[SIII]$_{33.5 \, {\rm {\mu m}}}$, and [SiII]. In general, [NeII] and
[SiII] are the strongest emission lines, and they are detected in
$98\%$ and $89\%$ of the targets, in the SH and LH spectra,
respectively.  The [SIII]$_{33.5 \, {\rm {\mu m}}}$ is most often seen
in the LH spectra ($91\%$ of the galaxies).  As we will discuss below,
some of the sources exhibit emission lines with asymmetric profiles or
relative velocity shifts between lines.  A number of hydrogen
molecular lines are also detected in the GOALS sources, but they will
be the subject of another paper reporting on the properties of the
warm molecular gas in LIRGs (Petric et al., Stierwalt et al., in prep.).

\begin{figure}
  \begin{center}
    \includegraphics[angle=0,scale=0.5,trim=0 0 0 50]{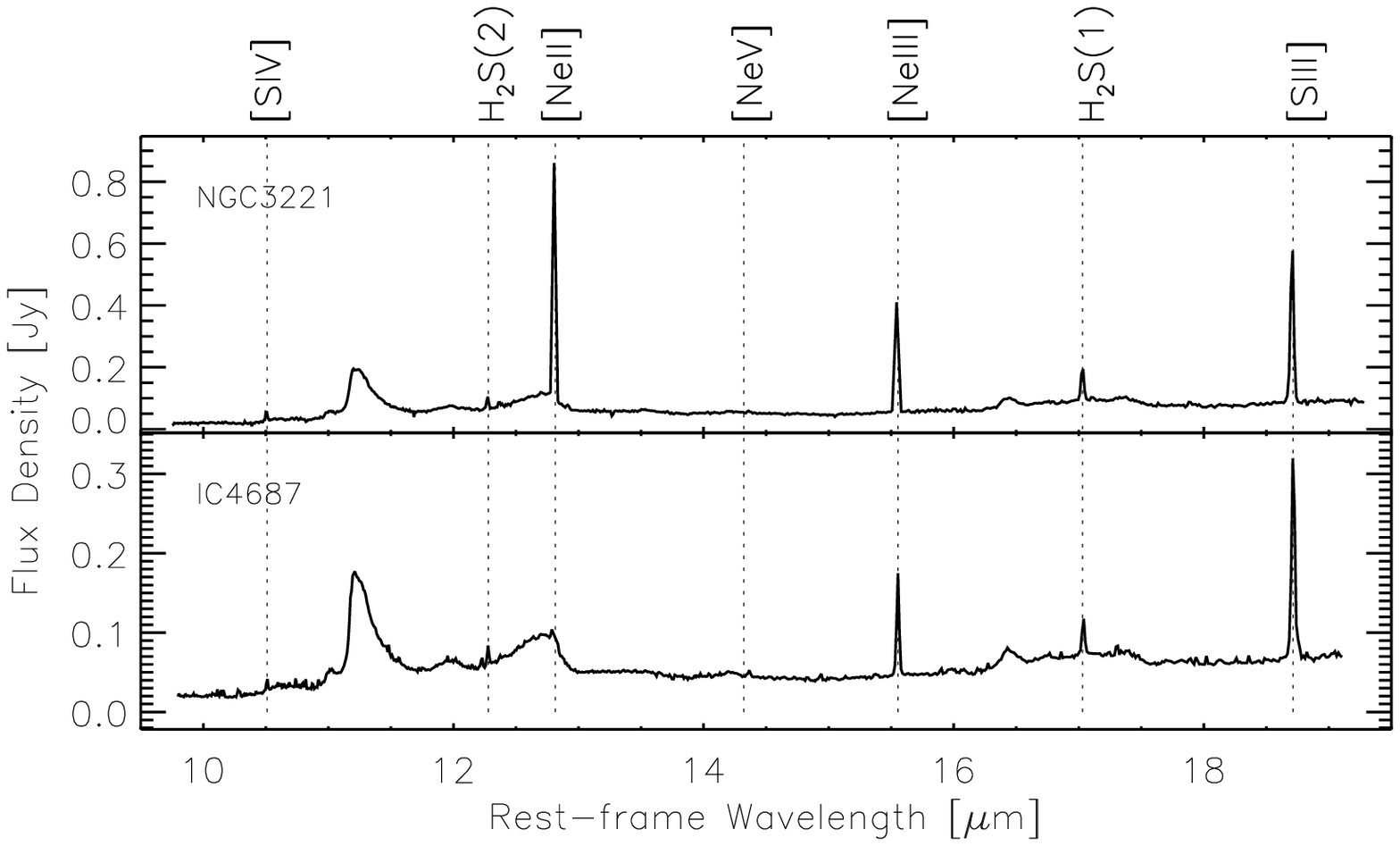}
    \includegraphics[angle=0,scale=0.5,trim=0 0 0 50]{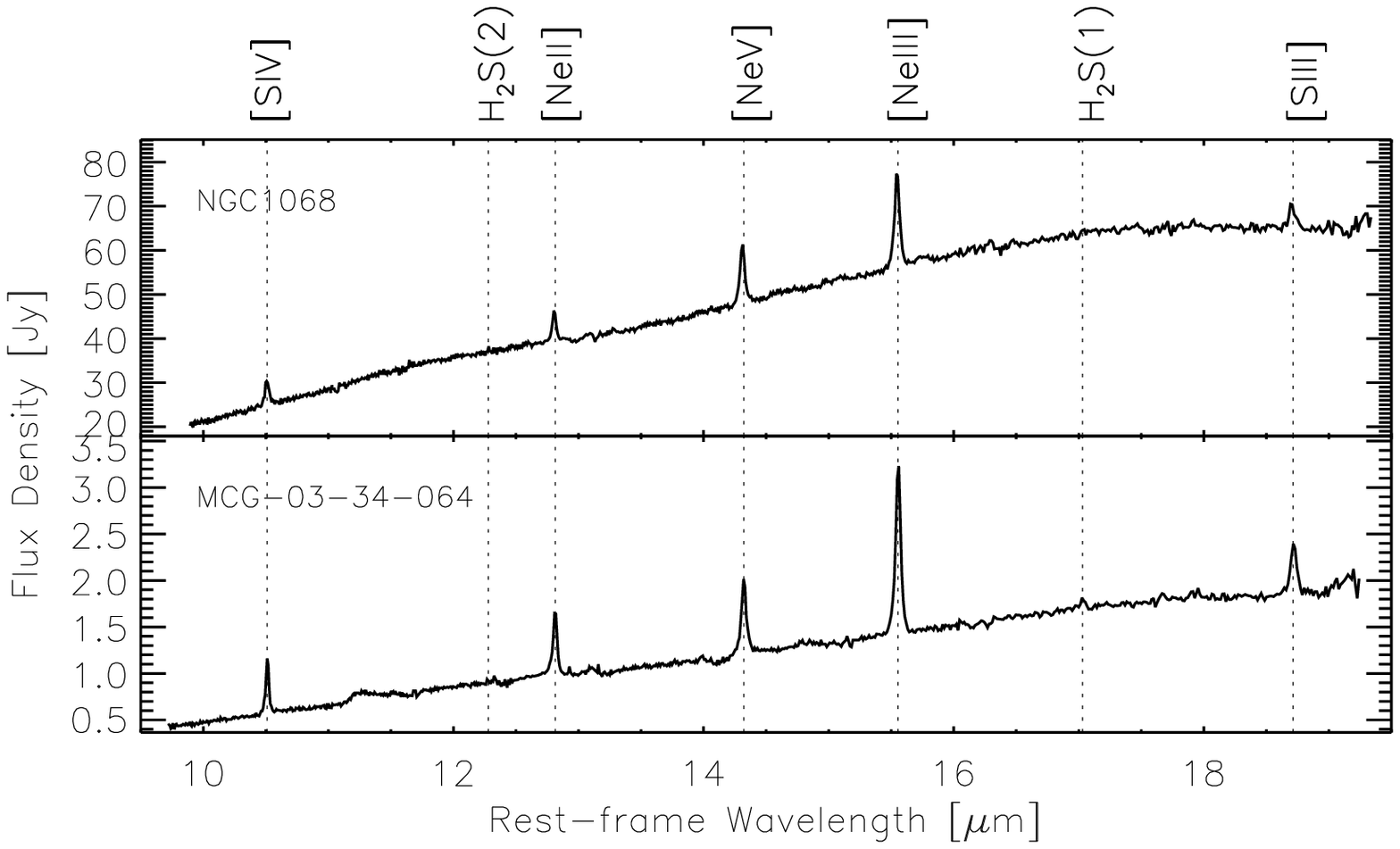}
    \caption{ Representative Spitzer IRS spectra
      of starburst (top) and AGN (bottom) dominated GOALS sources. Note the strong, 
      high-ionization lines of [NeV] and [SIV] in the bottom panels, along with the relatively weak 
      PAH emission at $11.3\mu$m. NGC~3221
      and IC~4687 have $6.2\,{\rm \mu m}$ PAH EQWs of $0.75\,{\rm \mu
        m}$ and $0.73\,{\rm \mu m}$, respectively. NGC~1068 and
      MCG-03-34-064 have [NeV]/[NeII] ratios of 1.96 and 1.10,
      respectively. }
    \label{fig:hires_spec}
  \end{center}
\end{figure}

A detailed analysis of various properties of the GOALS sample revealed
that $\sim 16-18\%$ of the sources show evidence of a central active
galactic nucleus \citep[AGN; see][]{Petr11}.  In this paper, we use
the $6.2 \mathrm{\mu m}$ polycyclic aromatic hydrocarbon (PAH)
equivalent width (EQW) and the [NeV] emission line flux to classify
starburst and AGN dominated sources.  While the [NeV] line fluxes are
reported in this paper, the EQWs are published in Table 1 of
\cite{Stie13a}.  Purely star-forming or starburst galaxies usually
have $6.2 \mathrm{\mu m}$ PAH EQW values in the range of $0.5-0.7 \,
\mathrm{\mu m}$ \citep{Bran06}.  A small $6.2 \mathrm{\mu m}$ PAH EQW
is often taken as an indicator of an AGN, since AGN typically have an
excess of hot dust which drives down the measured PAH EQW.  The
emission line at $14.3 \, \mathrm{\mu m}$ produced by $\rm Ne^{4+}$
corresponds to an ionization potential of $97.1 {\rm eV}$, which is
generally too high to be produced by young massive stars.  Therefore,
its presence in the composite spectrum of a galactic nucleus also can
signal the presence of an AGN.  Typically, the line flux ratio of the
[NeV]$_{14.3\,{\rm \mu m}}$ to [NeII]$_{12.8\,{\rm \mu m}}$ is used as
a diagnostic of the radiation field, because the ionization potential
of $\rm Ne^{+}$ is very low ($21.6 {\rm eV}$) compared with $\rm
Ne^{4+}$ and it is abundantly produced in HII regions.
In this paper, for the purposes of classification, sources are
considered to be AGN dominated in the mid-infrared if they have $6.2
\mathrm{\mu m}$ PAH EQW $\leq 0.3 \mathrm{\mu m}$ {\it or}
[NeV]/[NeII] $\geq 0.1$.  There are 57 GOALS nuclei in this category.
Note that this is slightly higher than the number reported in
\cite{Petr11}, because we are most interested here in understanding
the nature of the dominant starburst population, and have attempted to
be a bit more conservative in defining the starburst population among
the GOALS sources, and therefore excluding a few more composite
sources than \cite{Petr11}. Two sets of representative starburst and
AGN dominated spectra are shown in Figure~\ref{fig:hires_spec}.

\subsection{Comparisons with Starburst and Continuous Star Formation
  Models: Densities, Ages, Metallicities, and Ionization
  Parameters}\label{sec:SB_grid}


Ratios of emission lines arising from the same ionized species are
strong diagnostics of the interstellar medium (ISM). When these lines
originate from different transitions but of the same excitation level,
such as [SIII]$_{33.5\mathrm{\mu m}}$ and [SIII]$_{18.7\mathrm{\mu
    m}}$, their ratio is a good tracer of the gas density.  A total of 183
sources of our sample ($74\%$) have both of these lines detected in
LH. The [SIII]$_{33.5\mathrm{\mu m}}$ to [SIII]$_{18.7\mathrm{\mu
    m}({\rm LH})}$ line flux ratio ranges from $0.44$ to $4.18$, with
mean and median values of 1.66 and 1.61, respectively.  There are 112
GOALS nuclei with electron densities, as measured from the [SIII] line
flux ratios, more than $2\sigma$ above the low density limit
\citep[$\sim 100\,{\rm cm^{-3}}$ corresponding to an ${\rm [SIII]}$
33.5/18.7 flux ratio of 2.22 at $10^4$~K,][]{Drai10}. Of these, 91 are
starburst dominated. There are 69 sources that have their [SIII]
ratios consistent with gas at the low density limit. The mean and
median electron densities found for the GOALS sources with electron
densities above the low density limit are $360\,{\rm cm^{-3}}$ and
$300\,{\rm cm^{-3}}$, respectively. If we only consider the starburst
galaxies among these, then the mean and median values are $340\,{\rm
  cm^{-3}}$ and $300\,{\rm cm^{-3}}$, respectively. There are 2
galaxies with the [SIII] line flux ratios more than $2\sigma$ above
2.22 (\object{NGC~3221} and \object{NGC~7771}\_N), implying electron
densities below the low density limit. These sources may have
significant amounts of line-emitting gas at lower temperatures
\citep[$3000 - 7000$~K, see][]{Drai10} or they may have extended
[SIII] emission which preferentially raises the $33 \, {\rm \mu m}$
[SIII] line flux due to light scattering into the slit at longer
wavelengths. Throughout this paper, we employ a model with a density
of $n_e=100\,\mathrm{cm^{-3}}$. This is the closest available value to
our estimated electron densities and is sufficiently accurate given
the coarse resolution of the models (the next higher density is
$n_e=10^{4}\,\mathrm{cm^{-3}}$) and the slow change in the derived
properties with density at these values.


In order to reproduce the widest range of model parameters and avoid
degeneracies in the emission line ratios, we have selected the
[SIV]/[NeII] and [NeIII]/[NeII] line ratios as the primary diagnostics for
this study. Out of the 244 nuclei in our LIRG sample, [NeII], [NeIII],
and [SIV] emission lines are detected in 239, 234, and 103 objects,
respectively.  We note that because the SH and LH slit sizes are
different and many LIRGs are resolved in the SH wavelength range, we
try to avoid direct comparisons between emission lines detected in
different slits. When the emission lines from different apertures are
compared, we explicitly note that in the text or the figure captions.

The ionization potentials of the $\rm S^{3+}$, $\rm Ne^{+}$, and $\rm
Ne^{2+}$ emission lines are $34.8 {\rm eV}$, $21.6 {\rm eV}$, and
$41.0 {\rm eV}$, respectively. Even though the [SIV] emission line at
$10.5 \, {\rm \mu m}$ falls near the edge of $9.7\,{\rm \mu m}$
silicate absorption feature, we do not correct the measured line
fluxes for extinction in this paper. The range of the optical depth of
this absorption feature ($\tau_{9.7{\rm \mu m}}$) for the GOALS
sources spans $0.0 \leq \tau_{9.7{\rm \mu m}} \leq 10.8$
\citep{Stie13a}.  The apparent silicate optical depth
$\tau_{9.7{\rm \mu m}}$ can be converted to the extinction at the
$V$-band as $A_V = 19 \times \tau_{9.7{\rm \mu m}}$ \citep{Roch84}.
Although this does not directly trace the extinction in the gas, if we
were to use $\tau_{9.7{\rm \mu m}}$ and the equation (4) in
\cite{Smit07} to correct the [SIV], [NeII], and [NeIII] emission lines
for dust extinction, the [SIV]/[NeII] ([NeIII]/[NeII]) ratios would
increase (decrease) on the order of only 4\% (3\%). Therefore, even if
the silicate dust absorption seen in the GOALS sources were applied,
the relative extinction between the mid-infrared emission lines and
its effect on line ratio diagnostics would be very small.

\begin{figure}
  \begin{center}
    \includegraphics[angle=0,scale=0.5]{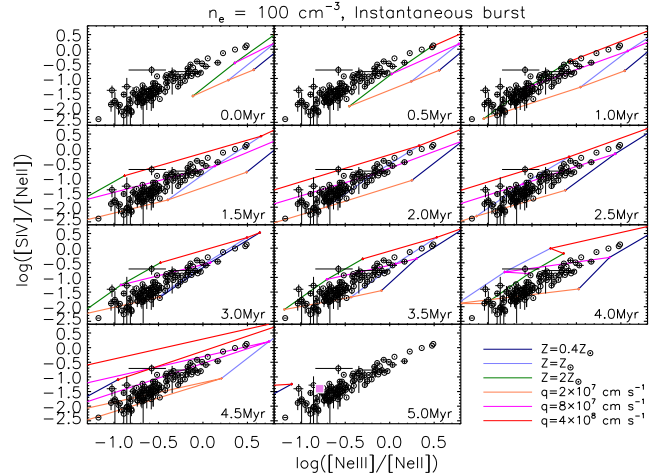}
    \caption { [SIV]/[NeII] vs. [NeIII]/[NeII] line flux ratio plots for the GOALS
      sample compared with an instantaneous starburst model assuming
      an electron density of $n_e=10^2 \, \mathrm{cm^{-3}}$.  The age
      evolution of the line ratios is shown from the top left to the
      bottom right panels in 0.5\,Myr step.  The parameters of the
      model are age (indicated in each panel), metallicity (dark blue:
      $Z=0.4Z_\odot$, light blue: $Z=Z_\odot$, green: $Z=2Z_\odot$),
      and ionization parameter (orange: $q=2\times10^7 \, \mathrm{cm
        \, s^{-1}}$, magenta: $q=8\times10^7 \, \mathrm{cm \, s^{-1}}$,
      red: $q=4\times10^8 \, \mathrm{cm \, s^{-1}}$). The pink filled
      square in the last panel denotes the median values of
      $\log{\rm ([SIV]/[NeII])}$ and $\log{\rm ([NeIII]/[NeII])}$, for
      the entire sample, including those sources with solid Ne line detections
       but no [SIV], and therefore it is not located at the center of the points in the 
        plot (which only include sources with solid [SIV] detections). }
    \label{fig:SIV_NeII_all}
  \end{center}
\end{figure}

In Figure~\ref{fig:SIV_NeII_all}, we present the [SIV]/[NeII]
vs. [NeIII]/[NeII] as a function of age for an instantaneous starburst
for the entire GOALS sample.  The vertical lines and the horizontal
lines denote metallicities and ionization parameters, respectively.
The median values of $\log{\rm ([SIV]/[NeII])}$ and $\log{\rm
  ([NeIII]/[NeII])}$ are $-1.40 \pm 0.06$ and $-0.79 \pm 0.03$,
respectively.

We have selected the [SIV]/[NeII] and [NeIII]/[NeII] line ratios as
primary diagnostics. We can see from the model grids that both
[SIV]/[NeII] and [NeIII]/[NeII] are sensitive to metallicity and
ionization parameter, but this combination of the line ratios let us
avoid degeneracies and maximizes the number of the sources in a
diagram for better statistical analysis. However, if we use
[SIV]/[SIII]$_{18.7\mathrm{\mu m (SH)}}$ instead of [SIV]/[NeII], the
model grids would move toward higher ionization (upward in
Figure~\ref{fig:SIV_NeII_all}).

\begin{figure}
  \begin{center}
    \includegraphics[angle=0,scale=0.5]{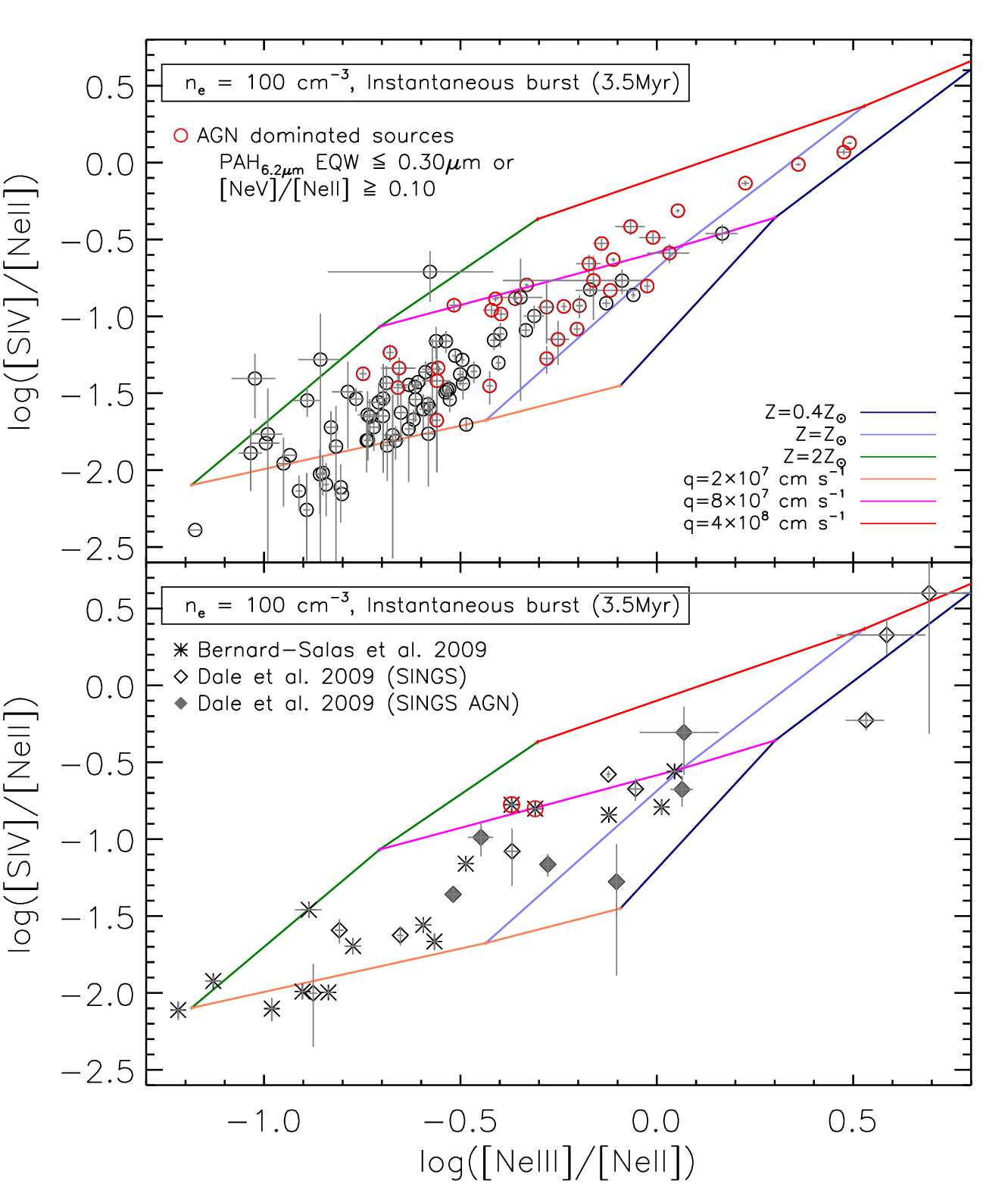}
    \caption { [Top panel] The same as Figure~\ref{fig:SIV_NeII_all},
      but only the 3.5\,Myr starburst model is shown against the GOALS data.  Red circles
      represent the AGN dominated sources with $6.2 \mathrm{\mu m}$
      PAH EQW $\leq 0.3 \mathrm{\mu m}$ or [NeV]/[NeII] $\geq 0.1$.
      [Bottom panel] The same as the top panel, but the data points
      are from \cite{Bern09} and \cite{Dale09} shown as the asterisks
      and the diamonds, respectively. Among the sources of
      \cite{Dale09}, the AGN sources, classified by \cite{Mous10}
      based on optical spectra, are indicated by the filled diamonds.}
    \label{fig:SIV_NeII_one}
  \end{center}
\end{figure}

\begin{figure}
  \begin{center}
    \includegraphics[angle=0,scale=0.5]{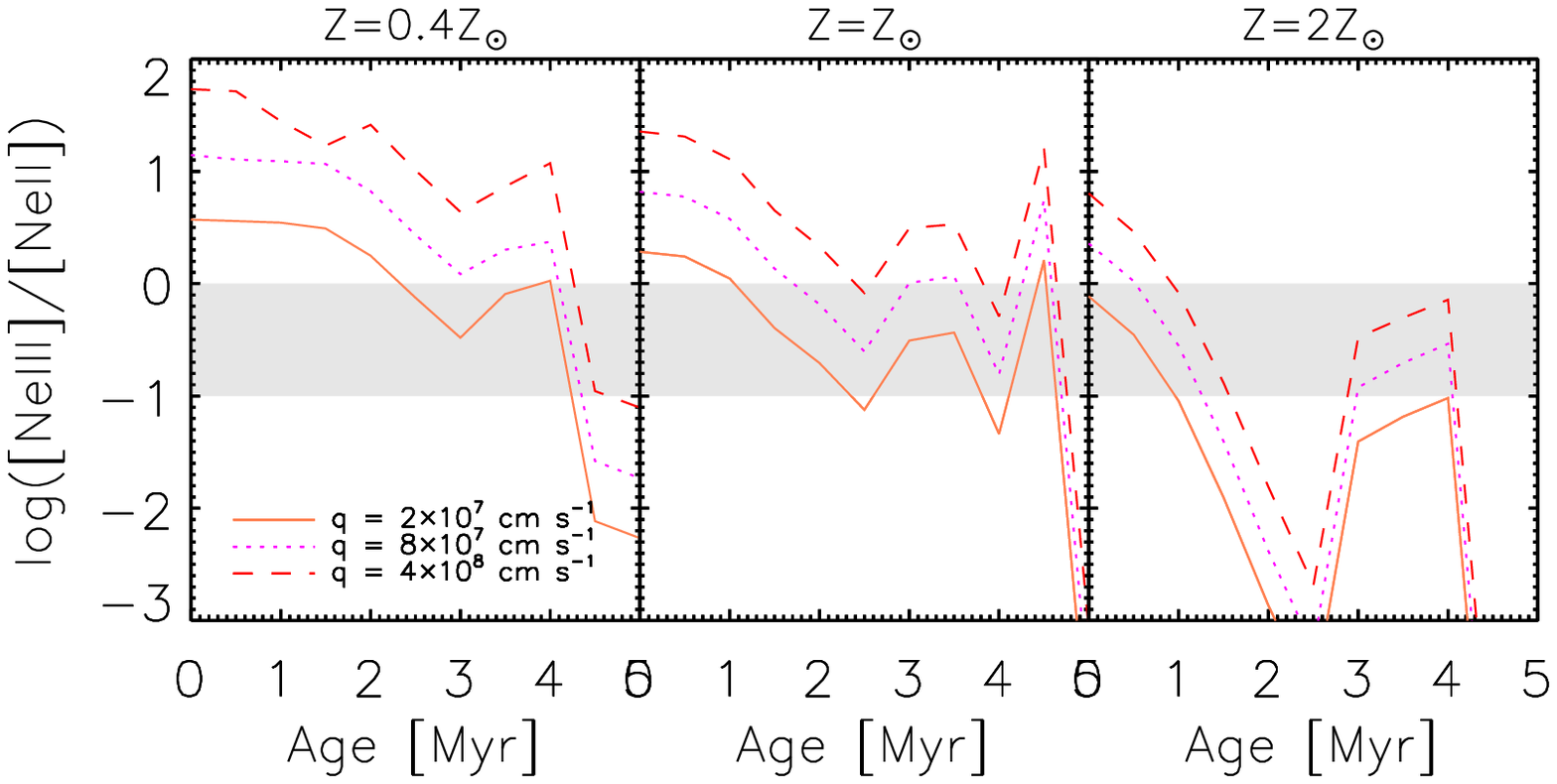}
    \includegraphics[angle=0,scale=0.5]{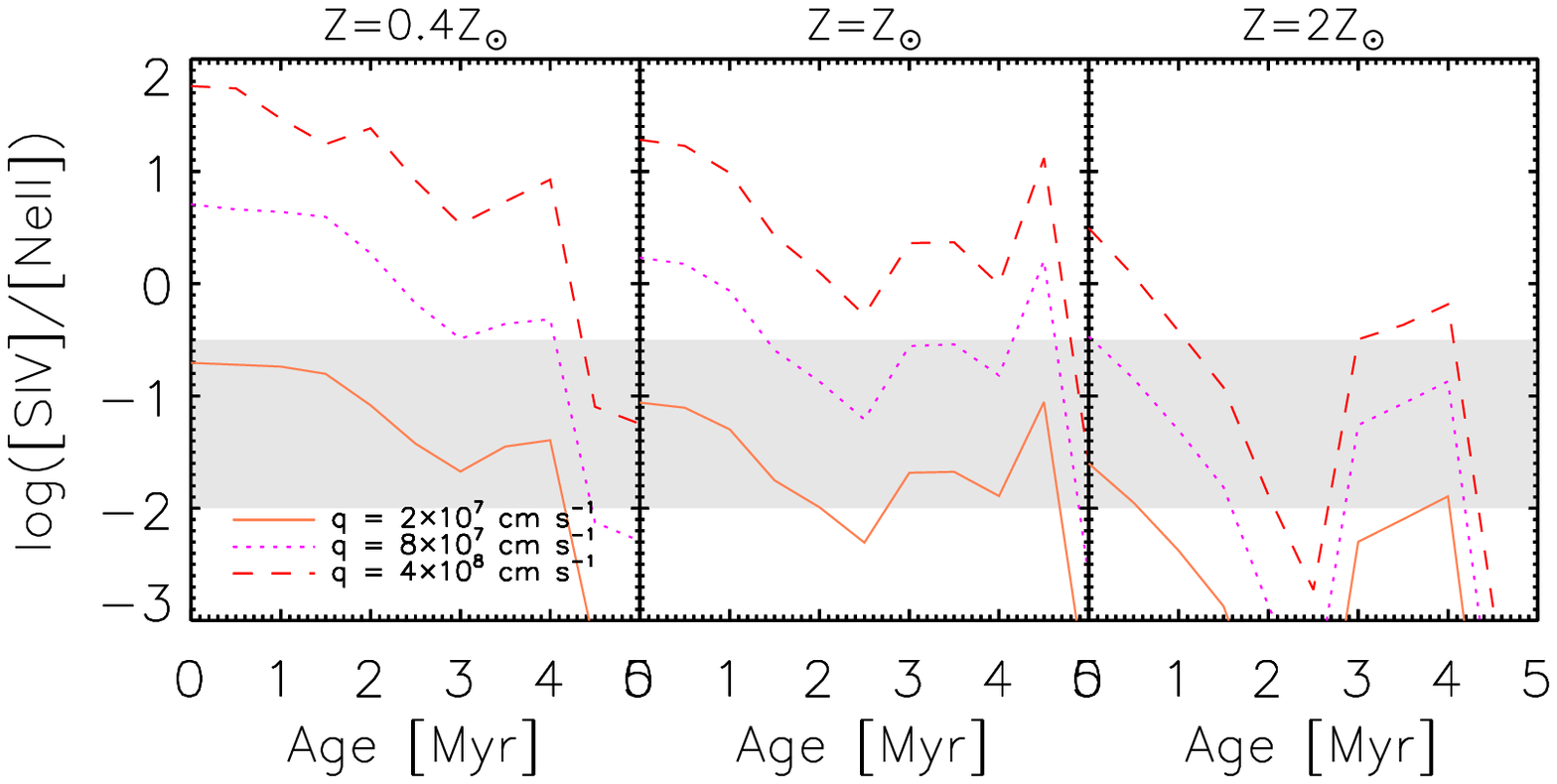}
    \caption { Age evolution of the ratios of
      [NeIII]/[NeII] (top) and [SIV]/[NeII] (bottom) for metallicities
      of $Z=0.4Z_\odot$, $Z=Z_\odot$, and $Z=2Z_\odot$ from left to
      right.  The Geneva Standard mass-loss evolutionary model with an
      electron density of $n_e=10^2 \, \mathrm{cm^{-3}}$ is used.  The
      orange solid, magenta dotted, and red dashed lines indicate the
      ionization parameters of $q=2\times10^7 \, \mathrm{cm \, s^{-1}}$,
      $q=8\times10^7 \, \mathrm{cm \, s^{-1}}$, and $q=4\times10^8
      \, \mathrm{cm \, s^{-1}}$, respectively.  The gray regions
      represent the range seen among the starburst-dominated GOALS sources.
    }
    \label{fig:temp_age}
  \end{center}
\end{figure}

With the assumption of the instantaneous star formation history, the
Pauldrach/Hillier \& Miller atmospheres, the standard Geneva stellar
evolution track, and the Salpeter IMF (see also
\S~\ref{subsec:phot_model}), the majority of the GOALS sources have
line flux ratios in agreement with the starburst models with ages
between $1-4.5$~Myr. Note that we are showing all the GOALS data in
this figure, including those sources that are thought to harbour AGN
which dominate their mid-infrared emission.  These AGN sources are
explicitly labeled in Figure~\ref{fig:SIV_NeII_one}, and they populate
the tail of the distribution of sources towards the upper right, but
do not change our estimate of the typical starburst age because they
run parallel to the age lines here.  In this age range there is a
dramatic evolution indicated by the line flux ratios; massive stars
are disappearing while Wolf-Rayet (W-R) stars start to dominate the
spectra \citep{Leve10}. This evolutionary path can be seen in
Figure~\ref{fig:temp_age}, where the ratios of [NeIII]/[NeII] and
[SIV]/[NeII] against the age are shown.  The regions filled with the
gray color correspond to the ranges of line ratios of only the
starburst GOALS sources.  In all of the metallicity ranges, the line
ratios decrease from $0$ to $\sim 3$~Myr due to the death of O/B-type
stars. After this phase, the line ratios rise rapidly because of the
emergence of the W-R stars.  The implied ionization parameters range
from $q=2-40\times10^7 \, \mathrm{cm \, s^{-1}}$ for the entire
sample, but only $q=2-8\times10^7 \, \mathrm{cm \, s^{-1}}$ for the
starburst dominated sources (see Figure~\ref{fig:SIV_NeII_one}).

The metallicities of the sources are constrained to be $1 \lesssim
Z\,[Z_\odot] \lesssim 2$ (Figure~\ref{fig:SIV_NeII_all}) showing that
LIRGs are usually metal-rich systems. For 22 GOALS systems, whose
metallicities were measure by \cite{Rupk08} via optical spectra, we
find a general agreement on the derived oxygen abundances when the
\cite{Trem04} calibration is used. These results are also in agreement
with recent studies of the metallicity gradients of LIRGs showing
overall metal rich systems using the \cite{Kewl02} calibration, which
differs from the \cite{Trem04} calibration by $< 0.1$~dex
\citep{Rupk10b,Rich12}.

\begin{figure}
  \begin{center}
    \includegraphics[angle=0,scale=0.5]{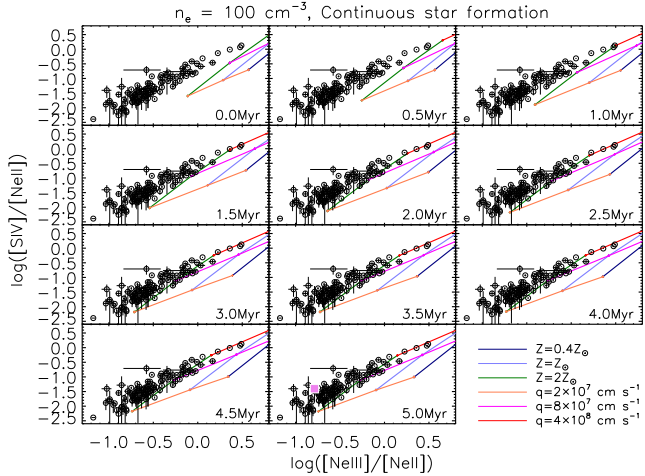}
    \includegraphics[angle=0,scale=0.4]{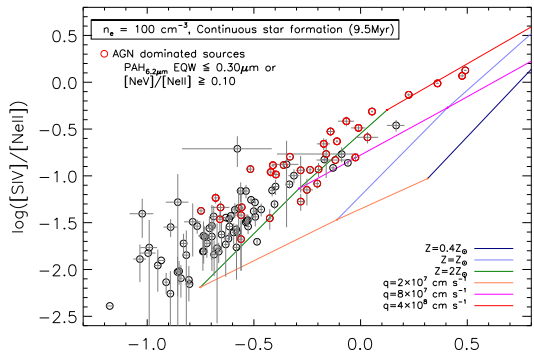}
    \caption { [Top] The same as in Figure~\ref{fig:SIV_NeII_all}, but for a
      comparison with the continuous star formation model.  [Bottom]
      The same as the top panel, but for the oldest age, 9.5\,Myr, in the model. }
    \label{fig:SIV_NeII_all_cont}
  \end{center}
\end{figure}

In the top panels of Figure~\ref{fig:SIV_NeII_all_cont}, we present
the same data as in Figure~\ref{fig:SIV_NeII_all}, but here we compare
the GOALS data to a model of continuous star formation. Most of the
sources have [NeIII]/[NeII] ratios which are too low to be reproduced
by the continuous star formation model, unless we invoke metallicities
greater than $2\,Z_\odot $ and ages above $5$~Myr.  Even then the
continuous star formation models can only fit a small fraction of the
sample.  This is not surprising given the fact that many of the LIRGs
are major mergers, which are modeled to have strong starburst peaks at closest passage
and final coalescence. For reference, in the bottom
panel of Figure~\ref{fig:SIV_NeII_all_cont} we also show the same plot
at the age of $9.5$~Myr (the oldest age allowed by the model). This is
long after the ionizing spectrum has reached a steady state so that
line ratios have become fully stabilized. For the remainder of this
paper we only consider the instantaneous starburst models when
comparing to the IRS data.

The relatively narrow range in starburst ages ($1-4$~Myr) we find for
the sample is expected, since young stars dominate the
ionization of the gas producing the emission lines visible in the IRS
data.  However, to place these ages into context, here we compare them
to other age estimates from the literature for LIRGs and ULIRGs.

Based on measurements of the EQW of the H$\beta$ absorption feature in
high resolution optical spectra in a sample of 25 nearby ULIRGs,
\cite{Soto10} find central starbursts surrounded by intermediate age
stellar populations indicating a rapid decrease in the star formation
rate $300-400$~Myr ago on radial scales of $5-10$~kpc. At the median
distance of our sample ($\sim 100$~Mpc), our IRS spectra cover the central
$2-5$~kpc of the GOALS sources.
While the H$\beta$ absorption EQW is most sensitive to A-type stars
which have longer lifetimes (a few Gyr), our emission line diagnostics
are most sensitive to young stars (O/B-type stars).  The stellar age
estimated by \cite{Soto10} traces the time since the most recent
episode of star formation, but the age that we estimate from the
mid-infrared emission lines probes the age of the most recent major
starburst. In addition, differences in ages could be caused by
differential extinction between the youngest and oldest stars in these
dusty nuclei, since some fraction of the young starburst probed with
our IRS data might be undetected in the visual.

\begin{figure}
  \begin{center}
    \includegraphics[angle=0,scale=0.5]{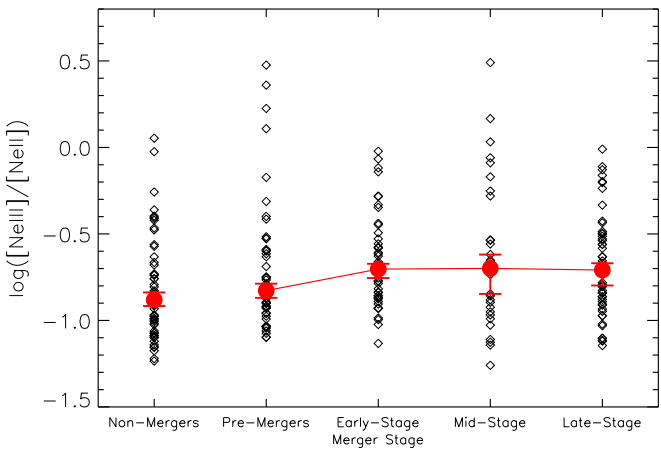}
    \caption{ The [NeIII]/[NeII] ratio as a function of the merger
      stage for GOALS sources. The red filled circles are the median values of the
      [NeIII]/[NeII] ratios within each merger stage. 
      All of the merger stages are classified by visual
      identification using the HST/ACS {\it B}- and {\it I}-bands
      imaging data for the objects with $\log{(L_{IR}/L_{\odot})} >
      11.4$ following \cite{Haan11a} and the IRAC imaging data for the
      rest of objects \citep{Stie13a}.  }
    \label{fig:Ne_Merger}
  \end{center}
\end{figure}

Since many of our LIRGs have experienced an interaction or merger, we
have compared our starburst ages to the dynamical timescales estimated
from imaging \citep{Haan11a} and $N$-body simulations \citep{Lotz08b}.
From a detailed study of the HST NICMOS imaging of GOALS sources with
$\log{(L_{IR}/L_{\odot})} > 11.4$, \cite{Haan11a} estimate a median
time until merger of $430$~Myr, with peaks at $\sim800$~Myr
(corresponding to first pass) and $< 300$~Myr (final coalescence).
Similarly, merger simulations \citep[e.g.,][]{Lotz08b} suggest
timescales of a few Gyr for a major merger, with star formation being
enhanced at first passage and final merger \citep{Miho94,HopkP08}.
Although late-stage mergers tend to have higher infrared luminosities
\citep[e.g.,][]{Veil02, Petr11}, we find no trend of the
fine-structure line ratios as a function of merger stages.  As an
example, in Figure~\ref{fig:Ne_Merger} we show the [NeIII]/[NeII] line
ratio as a function of merger stage. The merger stages are from Table
2 of \cite{Haan11a} and Table 1 of \cite{Stie13a}. This is likely due
to the fact that these ratios are always dominated by the youngest,
most massive stars, and hence they do not accurately probe
intermediate age populations formed earlier in the merger.

Super star clusters, which are often produced in great numbers during
a merger, can trace star formation over a longer time baseline,
provided extinction effects can be estimated
\citep{Whit95,Chie07}. For example, in the GOALS source
\object{NGC~2623}, star clusters are seen with ages consistent with
$\sim 1-100$~Myr \citep{Evan08}. In another luminous merger,
\object{II~Zw~096} (\object{CGCG~448-020}), there are two populations
of clusters with ages of $1-5$~Myr and $20-500$~Myr \citep{Inam10}.
Young and old cluster populations are also found in \object{NGC~7469}
\citep{Diaz07}. The younger star clusters ($1-5$~Myr) in
\object{II~Zw~096} are assumed to have been generated at the most
recent merger event.  While the dynamical age of \object{II~Zw~096} is
around 2~Gyr, it does host extremely young star clusters. Therefore,
while we are not sensitive with our mid-infrared ionized gas lines to
earlier bursts, the ages we are estimating are consistent with at
least some populations of extra-nuclear clusters in merging galaxies,
suggesting that we are probing the latest generation of stars in these
systems.

\subsubsection{The $3.5$~Myr Starburst Models}

In order to explore the detailed distribution of sources shown in
Figure~\ref{fig:SIV_NeII_all}, we expand the $3.5$~Myr photoionization
model grids in Figure~\ref{fig:SIV_NeII_one}.  For completeness in
presenting the entire distribution of GOALS sources, galaxies for
which an AGN may have a significant contribution to the mid-infrared
line flux (those with $6.2 \mathrm{\mu m}$ PAH EQW $\leq 0.3\,{\rm \mu
  m}$ or [NeV]/[NeII] $\geq 0.1$) are included in this figure, but
explicitly marked on the plot. While many of these sources may still
have their bolometric power dominated by star formation
\citep[see][]{Petr11}, there mid-infrared spectra are likely to have a
significant contribution from an AGN. Since the primary goal of this
paper is to explore the properties of the starbursts in LIRGs via
their mid-infrared fine-structure lines, we will exclude these
AGN-like sources from our analysis, even though we include them in
many plots after marking them where appropriate.  Most of the
starburst sources lie in the bottom left corner of the diagram with
median values of $\log{\rm([SIV]/[NeII])}$ and $\log{\rm
  ([NeIII]/[NeII])}$ of $-1.55 \pm 0.06$ and $-0.85 \pm 0.02$,
respectively.  They typically have ionization parameters between
$q=2-8\times10^7 \, \mathrm{cm \, s^{-1}}$ and metallicities between
$1-2\,Z_\odot$.


There are five highly ionized starburst sources in the top right
of Figure~\ref{fig:SIV_NeII_one} ($\log{\rm ([SIV]/[NeII])} > -1.0$
and $\log{\rm ([NeIII]/[NeII])} > -0.2$).  These sources, along with
their infrared luminosities $\log(L_{IR}/L_\odot)$ in parenthesis,
are: \object{IRAS~03582+6012}\_W ($11.43$), \object{IRAS~05129+5128}
($11.42$), \object{NGC~6052} ($11.09$),
\object{CGCG~448-020}\_W~\footnote{The nucleus of CGCG~448-020 we refer
  to is at J2000 (RA, Dec) $=$ (20:57:23.939, 17:07:36.098). The other
  nucleus (CGCG~448-020\_E) at (20:57:24.229, 17:07:40.110) in
  this system also has high [SIV]/[NeII] and [NeIII]/[NeII]
  ratios. However, although this nucleus seems to be dominated by an
  extreme starburst \citep{Inam10}, its $6.2\,{\rm \mu m}$ PAH EQW is
  $0.27\,{\rm \mu m}$, which is slightly smaller than our starburst
  threshold. } ($11.94$), and \object{ESO~148-IG002} ($12.06$).
The high ionization can be caused by young stellar population or metal
poor gas.  We found that the oxygen abundance of one of these
galaxies, \object{NGC~6052}, has been measured by \cite{Rupk08} to be
$8.79$ or $Z = 1.36 Z_{\odot}$, assuming the calibration of
\cite{Trem04}. This is consistent with the median value of their LIRG
sample.  While we do not have accurate metallicities for all the
sources with high excitation, it seems unlikely that metallicity
drives this effect in LIRGs. The $6.2\,{\rm \mu m}$ PAH EQWs of these
sources are large ($\gtrsim 0.5\,{\rm \mu m}$), with the exception of
\object{ESO~148-IG002}, which is on the boundary of our AGN
classification with a $6.2\,{\rm \mu m}$ PAH EQW of $0.31\,{\rm \mu
  m}$.  Its X-ray \citep{Iwas11} and optical (Rich et al. 2013)
spectra indicate that the southern nucleus is dominated by AGN.  The
corresponding [OIV]/[NeII] ratios of \object{IRAS~03582+6012}\_W,
\object{IRAS~05129+5128}, \object{NGC~6052}, and
\object{ESO~148-IG002} are $0.06 \pm 0.02$, $0.04 \pm 0.01$, $0.08 \pm
0.02$, and $0.47 \pm 0.08$ (\object{CGCG~448-020}\_W does not have a
[OIV] detection).  Note that, while these highly ionized starburst
LIRGs exist, their measured [NeIII]/[NeII], [SIV]/[NeII], and
[OIV]/[NeII] line flux ratios are still significantly lower than the
mean values found for the sources we believe to harbor a mid-infrared
dominant AGN.

\subsubsection{Comparisons with Other Starburst Samples}


A comparison of our sample with local starburst and star-forming
galaxies from \cite{Bern09} and \cite{Dale09} is presented at the
bottom panel of Figure~\ref{fig:SIV_NeII_one}.  The 24 targets in
\cite{Bern09} consist of 22 starburst galaxies and two galaxies with
evidence of AGN \citep[${\rm [NeV]/[NeII]} \geq 0.1$ or PAH EQW $\leq
0.30\,{\rm \mu m}$ from][]{Bran06}. Since we require a detection in
the [SIV], [NeII], and [NeIII] lines, only 15 of the Bernard-Salas
sources appear in the panel~\footnote{In total there are 16 sources
  which have all of [SIV], [NeII], and [NeIII] detections. However,
  the one that we excluded here, NGC~4945, has a [SIV]/[NeII] ratio of
  $(1.2 \pm 0.3) \times 10^{-3}$ and lies outside the plotted y-axis
  range. For reference, its [NeIII]/[NeII] ratio is $0.12 \pm 0.03$.}.
Seven of these sources (\object{NGC~1365}, \object{NGC~1614},
\object{NGC~2146}, \object{NGC~2623}, \object{NGC~3256},
\object{NGC~4194}, and \object{NGC5256\_S}) are also in the GOALS
sample, and the comparison of the measured line fluxes agrees in
nearly all cases to within $2-10\%$.  These objects have an infrared
luminosity $\log(L_{IR}/L_\odot)$ between $9.75-11.6$ with a median of
$10.7$.  Also, a subset of 15 out the 67 objects of \cite{Dale09}
appear in our figure due to a lack of large numbers of [SIV]
detections among the SINGS sources (Spitzer Infrared Nearby Galaxies
Survey) sources.  The SINGS galaxies cover a range of Hubble types and
infrared luminosities from $<10^{7}\,L_{\odot}$ to
$2\times10^{11}\,L_{\odot}$ \citep{Kenn03}.  Although primarily
star-forming and earl-type galaxies, the SINGS sample also includes
some sources with (weak) AGN -- these are marked in the figure to
avoid confusion.

There are 13 starbursts from \cite{Bern09} and six SINGS star-forming
galaxies that overlap with the location of the GOALS LIRGs in
Figure~\ref{fig:SIV_NeII_one}. 
A few SINGS galaxies with no AGN signature display extremely high
ionization, reaching $\log{\rm ([NeIII]/[NeII])} \sim 0.6$. Their
corresponding $\log{\rm ([SIV]/[NeII])}$ ratios are also high. Two of
these, \object{NGC~1705} and \object{NGC~2915}, are classified as blue
compact dwarfs~\footnote{Four starburst galaxies with even more
  extreme mid-infrared line ratios have been identified among a lower
  IR luminosity local blue compact dwarf sample by Hao et
  al. (2009).}.  The oxygen abundances ($12+\log{\rm (O/H)}$) of
\object{IC~4710}, \object{NGC~1705}, and \object{NGC~2915} were found
to be $8.18$, $7.96 \pm 0.06$, and $7.94 \pm 0.13$ respectively
\citep{Mous10}.  These values are lower than the average LIRG
metallicity \citep[see Figure~2 in][]{Rupk08}. Because the
[NeIII]/[NeII] ratios depend on the metallicities, these three sources
and the sources which harbor weak AGN \citep[diagnosed by optical
spectra;][]{Mous10} are excluded from a following comparison with the
GOALS starburst LIRGs.  The rest of the SINGS objects have similar
[SIV]/[NeII] and [NeIII]/[NeII] ratios to the GOALS starbursts,
covering a range of $-2.0 \lesssim \log{\rm ([SIV]/[NeII])} \lesssim
-0.5$ and $-1.0 \lesssim \log{\rm ([NeIII]/[NeII])} \lesssim 0.0$.


Since the detection of the high-ionization, weak [SIV] emission line
greatly limits the comparison of the GOALS sources to the the bulk of
the \cite{Bern09} and \cite{Dale09} sources, we can focus on examining
the distribution of the [NeIII]/[NeII] line flux ratios among the
three datasets.  The GOALS starburst sources are highly concentrated
in the range of $-1.0 \lesssim \log{\rm ([NeIII]/[NeII])} \lesssim
-0.5$, similar to the starburst galaxies of \cite{Bern09}.  The SINGS
galaxies show a wider distribution while covering the same general
range in [NeIII]/[NeII].
The two-sample Kolmogorov-Smirnov (K-S) statistical tests between the
GOALS starburst sample and \cite{Bern09} and \cite{Dale09} yields
$p$-values of $0.91$ and $0.10$, respectively, suggesting that the
[NeIII]/[NeII] ratios of the GOALS starburst sample and the starburst
sample of \cite{Bern09} are consistent with being drawn from the same
parent distribution, with a K-S test confidence level of
$91\%$. Surprisingly, although the SINGS sources have more quiescent
star formation than the GOALS sources, their median [NeIII]/[NeII]
value is higher (0.22 for SINGS vs. 0.14 for GOALS).  The range of the
oxygen abundances of the SINGS star-forming nuclei with both [NeIII]
and [NeII] detections is $8.12-8.71$ \citep{Mous10} with the median of
$8.43$, while it is $7.32-8.84$ with the median of $8.13$ for the
GOALS starbursts \citep{Rupk08}, both calibrated with
\cite{Pily05}. Therefore the slightly higher [NeIII]/[NeII] flux
ratios seen in the SINGS nuclei are not caused by an overall lower
metallicity compared with the GOALS starbursts. The difference may
however, be related to the smaller projected area of the IRS nuclear
data for the SINGS galaxies, which may provide less contamination from
surrounding, lower excitation gas.  The median projected area of the
SINGS and GOALS galaxies are $1.13 \, {\rm kpc^2}$ and $13.1 \, {\rm
  kpc^2}$, respectively.


It is well known that the [NeIII]/[NeII] line ratio reflects the
hardness of the radiation field. Thus it allows us to examine the
upper mass cutoff of the initial mass function (IMF) at a given
metallicity, star formation history, and starburst age \citep{Thor00}.
Our GOALS starburst dominated sources have [NeIII]/[NeII] ratios in
the range of $-1.3 \lesssim \log{({\rm [NeIII]/[NeII]})} \lesssim
0.1$, which limits the upper mass cutoff to $\sim 30-100 \, {\rm
  M_\odot}$. This range extends toward lower masses than what was
found by \citet[$\sim 50-100 \, {\rm M_\odot}$]{Thor00} in their
starburst sample.

\subsection{Comparison with Shock Models: 
 Existence of Shocks and Shock Speeds}\label{sec:shock}

In addition to photoionization, shocks can also excite gas in LIRGs
\citep{Monr10,Rich11}.  In the following sections, we compare the IRS
line flux ratios to those predicted to arise from shocked gas, as well as
discuss the location of LIRGs with resolved or shifted emission lines
with respect to the model grids.

\subsubsection{Shock Ionization}

\begin{figure}
  \begin{center}
    \includegraphics[angle=0,scale=0.5]{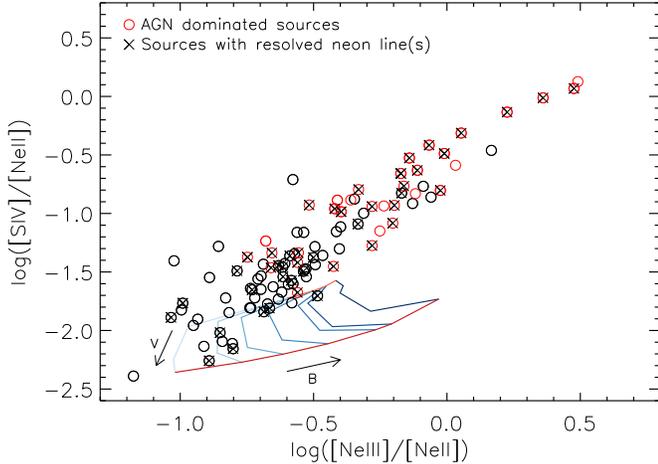}
    \caption{ The same line ratio diagram as
      Figure~\ref{fig:SIV_NeII_one}, but with the shock model grids
      \citep{Alle08} superposed. The AGN dominated sources (PAH$_{6.2
        \mathrm{\mu m}}$ EQW $\leq 0.3 \, \mathrm{\mu m}$ or
      [NeV]/[NeII] $\geq 1.0$) are indicated with the red circles. The
      black crosses ($\times$) denote the sources with at least one of
      the neon lines having a line width (FWHM) of $>
      600\,\mathrm{km\,s^{-1}}$ (see also
      Section~\ref{subsec:kinematics} and Figure~\ref{fig:Ne_width}
      for the details).  The parameters of shock velocities $V$ and
      the magnetic fields $B$ increase with the directions of the
      arrows, in steps of $100$, $200\,\mathrm{km\,s^{-1}}$ (the pink
      and red lines, respectively) and $1$, $5$, $10$, $20$, $40$,
      $50$, $100\,\mathrm{\mu G}$ (from the light blue to the dark
      blue lines), respectively. }
    \label{fig:SIV_NeII_shock}
  \end{center}
\end{figure}

In Figure~\ref{fig:SIV_NeII_shock}, we overlay the shock model grids
of \cite{Alle08}, with shock speeds of $100-200\,{\rm km\,s^{-1}}$ and
magnetic field strengths of $1-100\,{\rm \mu G}$, on the [SIV]/[NeII]
vs. [NeIII]/[NeII] emission line diagnostic diagram.  The model is
consistent with the lines ratios of 10 starburst
LIRGs. 
They are at the low-end of the [SIV]/[NeII] and
[NeIII]/[NeII] ratios seen in the GOALS sample.  While these galaxies
are consistent with shock ionization, very young starbursts with ages
around $1-2$~Myr (with an ionization parameter of $\sim 2 \times 10^7
\, {\rm cm \, s^{-1}}$; see Figure~\ref{fig:SIV_NeII_all}) can also
reproduce these line ratios.

\begin{figure}
  \begin{center}
    \includegraphics[angle=0,scale=0.5]{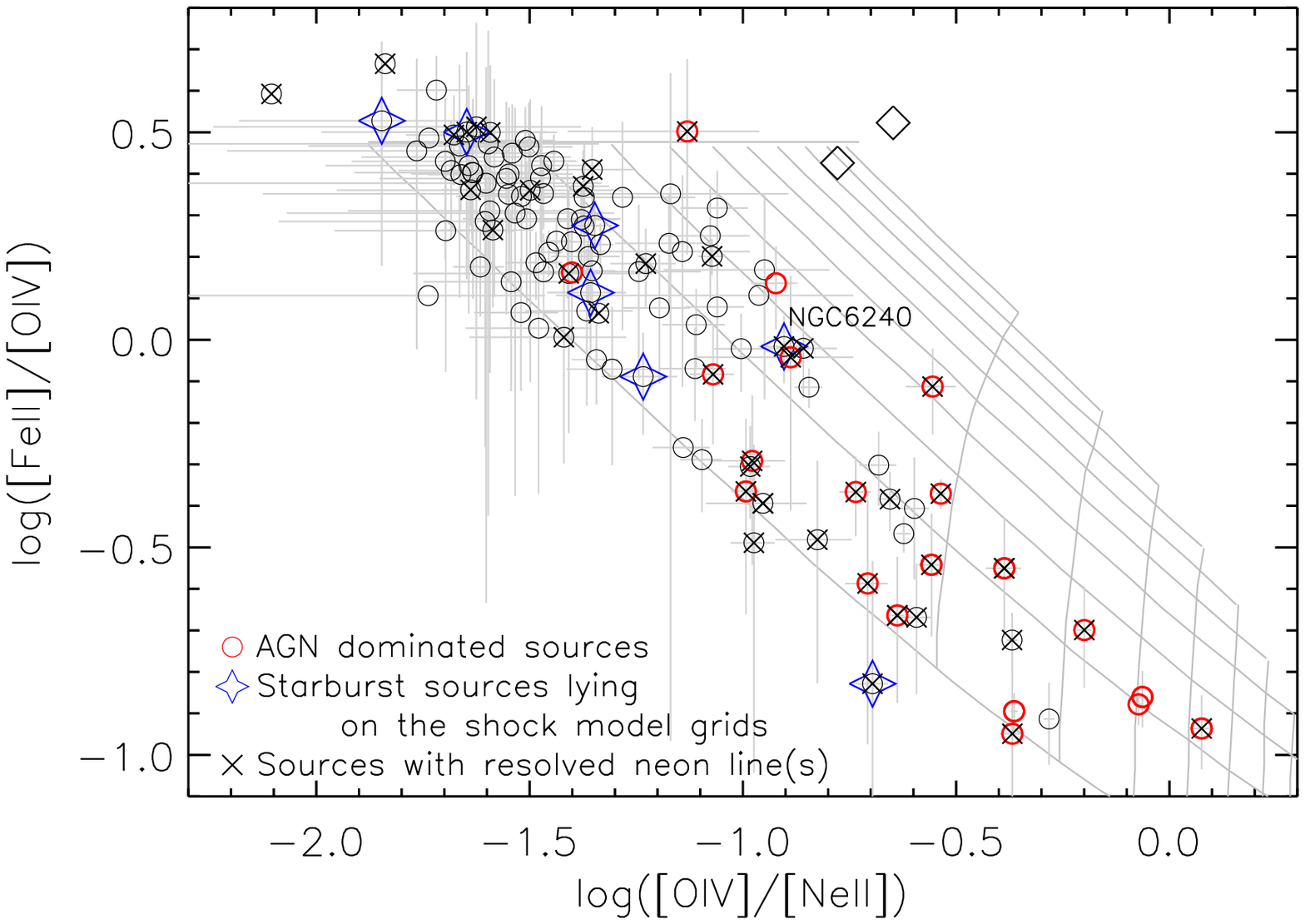}
    \caption { The [FeII]/[OIV]
      vs. [OIV]/[NeII] diagram for GOALS sources. The symbols are the same as
      Figure~\ref{fig:SIV_NeII_shock}, but here we also indicate the
      sources which lie on the shock model grids in
      Figure~\ref{fig:SIV_NeII_shock} as the blue stars. The two large
      diamonds at the top right are supernovae IC~443 and RCW~103
      \citep{Oliva99a,Oliva99b} for comparisons. The grids are simple
      mixing lines from \cite{Lutz03} which indicate contribution of
      shocks from supernova remnants or AGN (0\%, 10\%, 20\%,
      ...). 
    }
    \label{fig:FeII_OIV}
  \end{center}
\end{figure}

Due to strong depletion of Fe onto grains in the ISM \citep{Sava96},
the [FeII]/[OIV] together with the [OIV]/[NeII] line flux ratios can
be used to distinguish sources with strong shocks (which can destroy
grains and release Fe into the gas phase) from starbursts and AGN
\citep[e.g.,][]{Lutz03,Stur06}.  We plot the [FeII]/[OIV]
vs. [OIV]/[NeII] line flux ratios of the GOALS sources in
Figure~\ref{fig:FeII_OIV}.  Shock dominated sources (e.g., supernovae,
SNe) typically have both strong [FeII] and strong [OIV] emission, as
indicated by the positions of IC 443 and RCW 103 from
\cite{Oliva99a,Oliva99b}, and are clearly offset from the location of
starbursts (with weak [OIV] emission) and AGN (with very strong [OIV]
emission).  As expected, most of the AGN dominated LIRGs are found at
the lower right of Figure~\ref{fig:FeII_OIV}, while the starburst
dominated LIRGs mostly populate the upper left corner. However, there
are $\sim 10$ LIRGs with elevated [FeII]/[OIV] line flux ratios (for
their [OIV]/[NeII] line flux ratios) which depart from the locus
connecting the starburst dominated and AGN dominated LIRGs, towards the SNe.
Only one of these source, \object{NGC~6240}, which is known to exhibit
strong shock gas emission due to the interaction of two
counter-rotationg disks \citep[][and references therein]{Armu06},
falls on the shock model ionization grids discussed earlier
(Figure~\ref{fig:SIV_NeII_shock}).  The fact that the majority of the
LIRGs consistent with the \cite{Alle08} shock models do not show
enhanced [FeII] emission may suggest that these LIRGs are indeed
dominated by very young starbursts.  Similarly, the sources with
excess [FeII] emission may have some contribution from shocks to their
line ratios at the $< 10-30 \%$ level \citep[see the mixing lines in
the figure proposed by][]{Lutz03}.

\subsubsection{Gas Kinematics}\label{subsec:kinematics}

Despite the limited spectral resolution of IRS data for determining
gas kinematics, in some ULIRGs and AGN, IRS spectra have been
successfully used to study outflows and gas motion around black holes
\citep{Spoo09a,Spoo09b,Dasy11}.  Here we compare the widths and
velocity offsets between the [NeII], [NeIII], and [NeV] emission lines
to detect large-scale motions of the ionized gas, and associate these
with the ionization state of the gas.

We find that five sources show velocity shifts greater or equal to
$200\,{\rm km\,s^{-1}}$ in the [NeIII] or [NeV] emission lines
relative to the [NeII] emission line. All five sources are detected in
[NeV], indicating the presence of an AGN and only one
(\object{ESO~602-G025}; ${\rm [NeV]/[NeII]} = 0.05 \pm 0.01$ and
$6.2\,{\rm \mu m}$ PAH EQW $= 0.45 \pm 0.01 \,{\rm \mu m}$) is not
classified as an AGN dominated source.  In these galaxies at least,
the highly ionized gas is kinematically distinct from the bulk of the
emission, perhaps being located closer to the ionizing source and
participating in an outflow (or infall).
A total of 80 sources have at least one resolved neon emission line
with a FWHM $\geq 600\,{\rm km\,s^{-1}}$ corresponding to an intrinsic
line width of $\gtrsim 330\,{\rm km\,s^{-1}}$.  These sources cover a
wide range in the emission line flux ratios.  Approximately $60\%$
(48) of the sources with resolved neon lines are starbursts, with the
remaining 31 being AGN-dominated. All of the sources with large neon
line velocity shifts also have FWHM $\geq 600\,{\rm km\,s^{-1}}$. The
profiles of these emission lines and a table with the measured FWHM
are given in the Appendix.  In addition, there are 10 objects in our
sample exhibiting [OIV] emission lines shifted in velocity with
respect to [NeII] by more than $200 \, {\rm km\, s^{-1}}$.  There are
98 GOALS sources with resolved [OIV] emission lines, and 69 of these
are starburst-dominated.

The sources with resolved Ne emission lines are indicated in
Figure~\ref{fig:SIV_NeII_shock}.
Among the 10 sources lying on the shock model grids, five of them have
resolved neon emission lines, including \object{NGC~6240}.
One of these sources, \object{IRAS~13120-5453},
also shows evidence of outflows in the optical data: an asymmetric profile
in its H$\alpha$ emission and broad, blue-shifted wings in its Na~D
absorption profile (Rich et al. 2013, in preparation).
The rest of the LIRGs with resolved emission are either starbursts or
AGN. On average, the AGN have broader emission lines
than the starburst dominated LIRGs. The median FWHMs of [NeIII] for
starburst and AGN dominated sources are $540$ and $620\,{\rm km \,
  s^{-1}}$, respectively.

\begin{figure}
  \begin{center}
    \includegraphics[angle=0,scale=0.5]{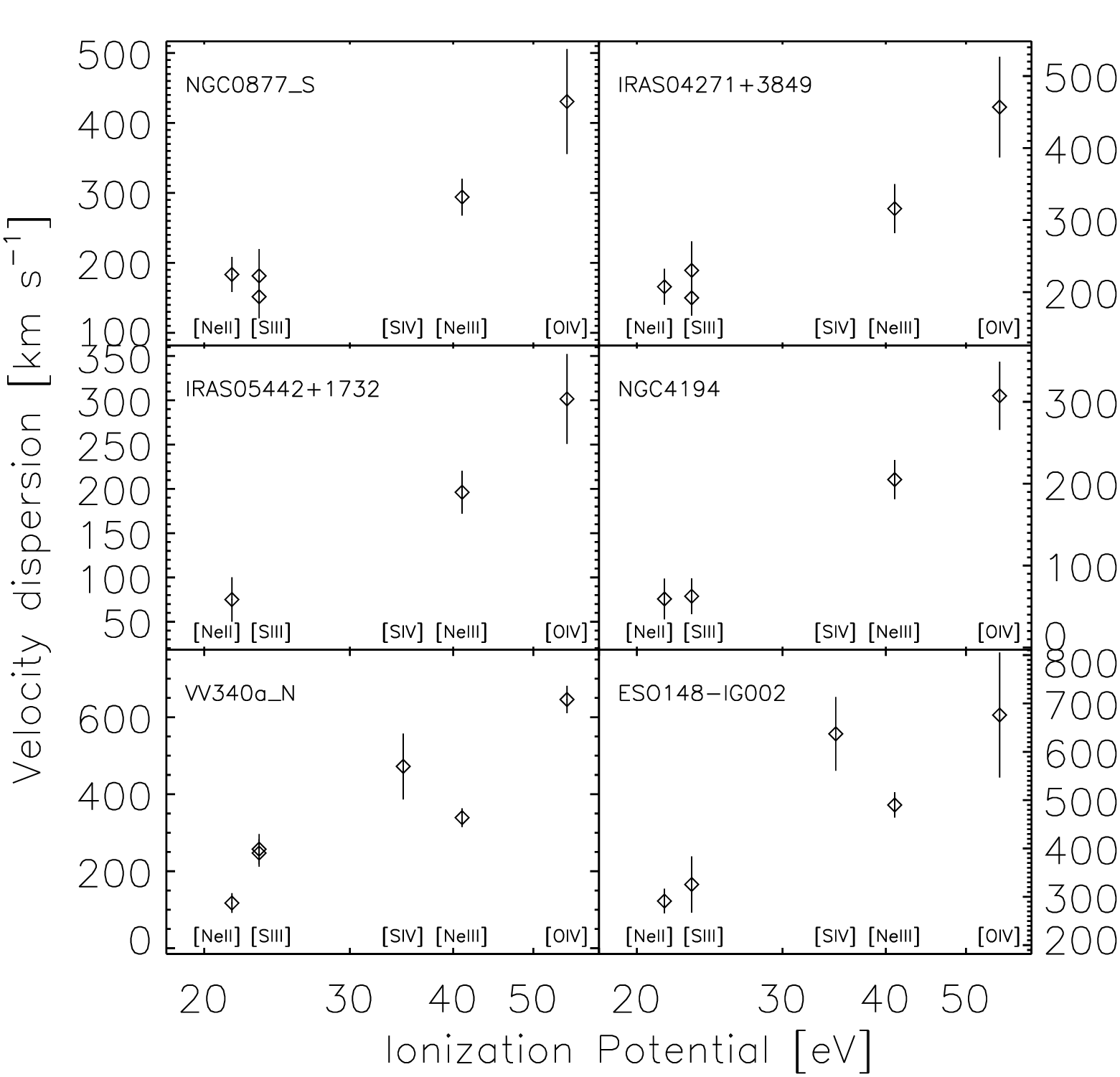}
    \includegraphics[angle=0,scale=0.5]{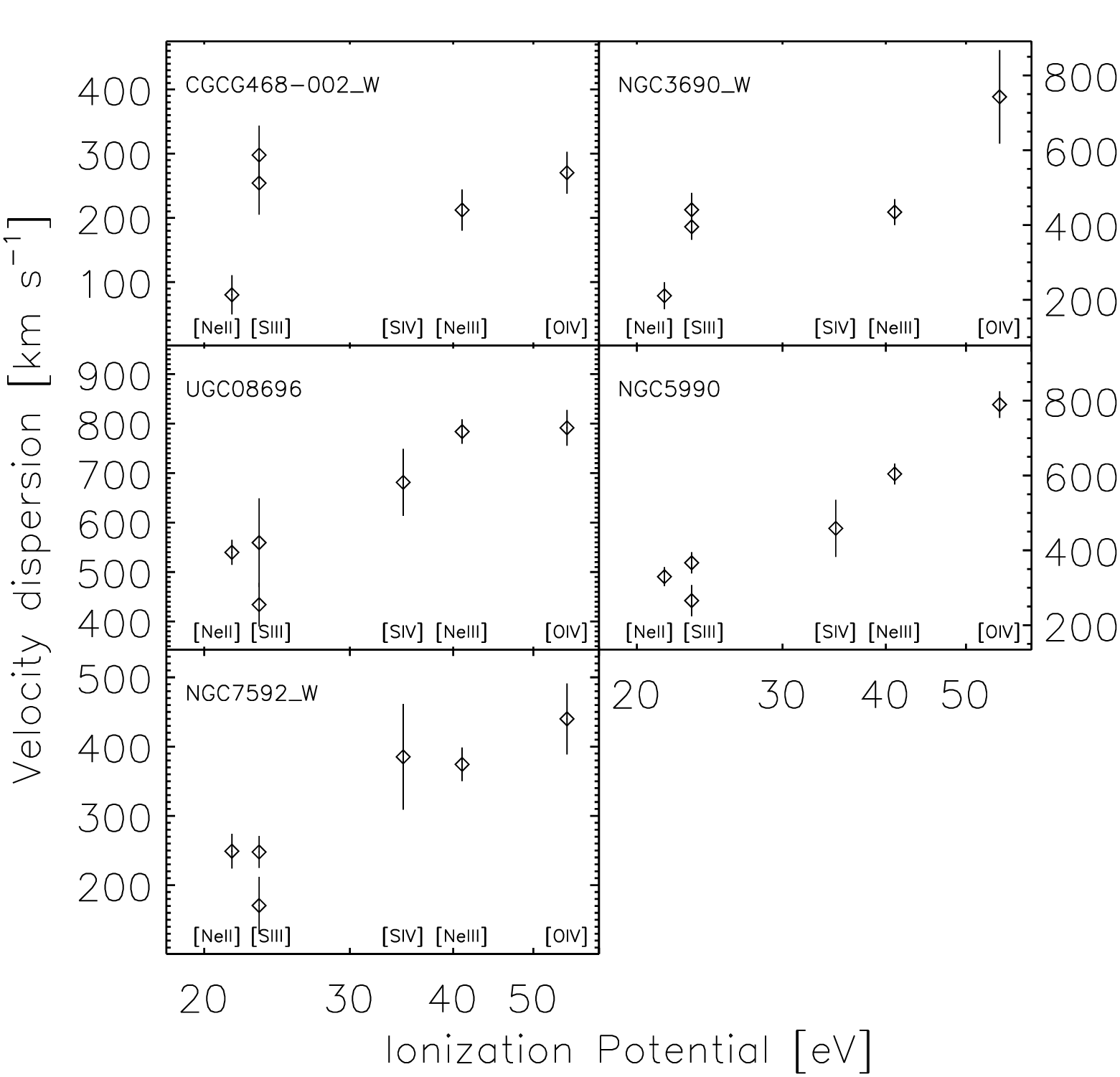}
    \caption { The line width (corrected for the intrinsic velocity
      dispersion) of each species as a function of the ionization
      potential. The presented six starburst (the top panel) and five AGN
      dominated sources (the bottom panel) show increasing velocity
      dispersion with ionization potential. }
    \label{fig:width_dens_poten}
  \end{center}
\end{figure}

Correspondence between the line widths and ionization may also be able
to shed light on the distribution of the ionized gas.  In its simplest
form, the emission line nebula around a compact energy source may show
a correlation of ionization potential and line width, reflecting the
energetic input of the source into the gas.  More ionized species,
which arise closer to the energy source, may display broader lines.
In the case of AGN, line widths, luminosities, and black hole masses
are also found to be correlated \citep{Dasy11}. There are 11 GOALS
sources (six starbursts and five AGN) that show a clear correlation of
line width and ionization potential.  All show increasing line width
for the higher ionization lines. The plots of line width
vs. ionization potential are shown in
Figure~\ref{fig:width_dens_poten}.  An additional eight LIRGs (six
starbursts and two AGN) may also show the same trends, but with less
than four detected lines in each source, we consider these as marginal
correlations.

\subsection{Line Fluxes as Star Formation Indicators}\label{subsec:LineSFR}

\begin{figure}
  \begin{center}
    \includegraphics[angle=0,scale=0.45]{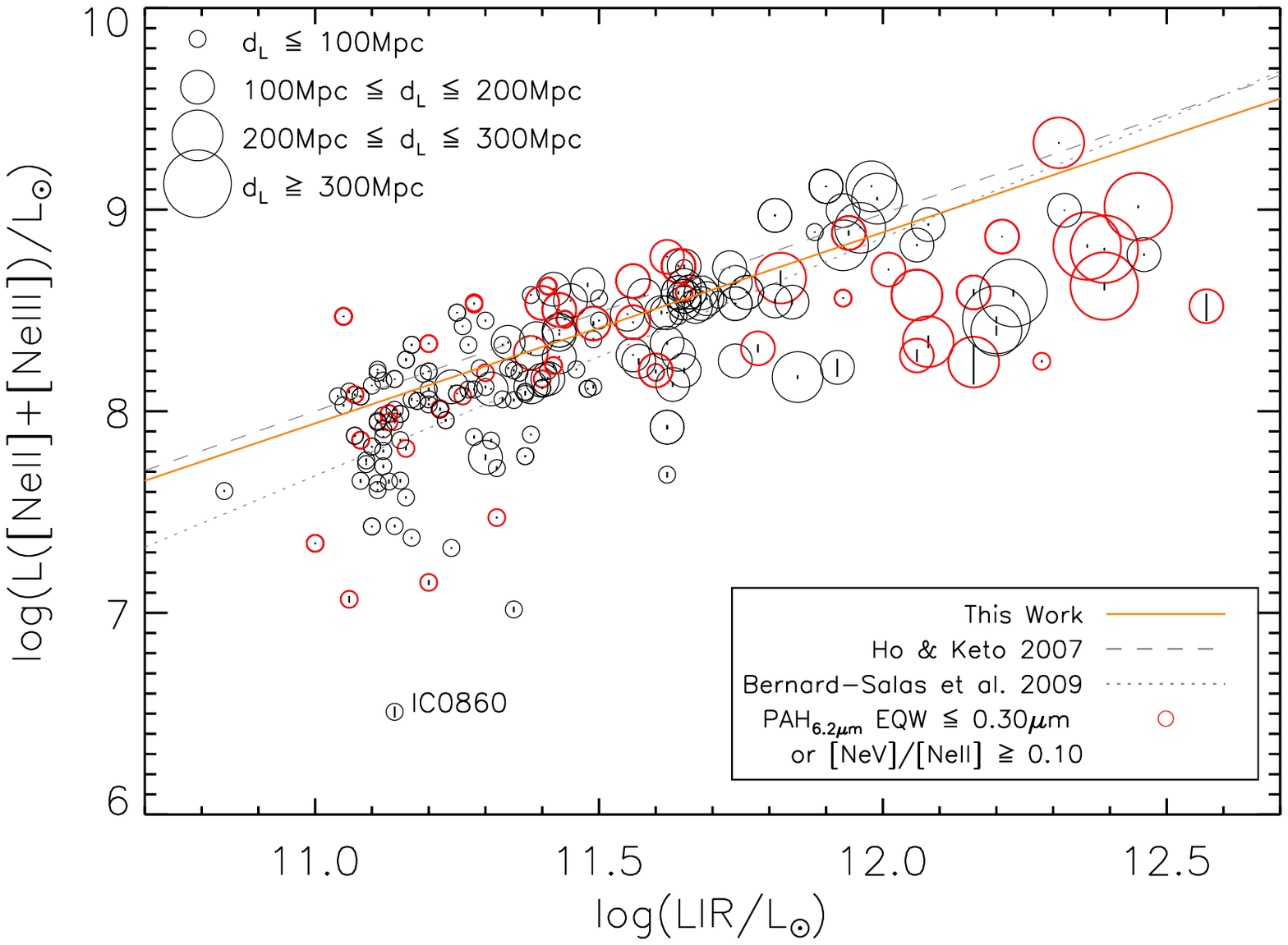}
    \includegraphics[angle=0,scale=0.45]{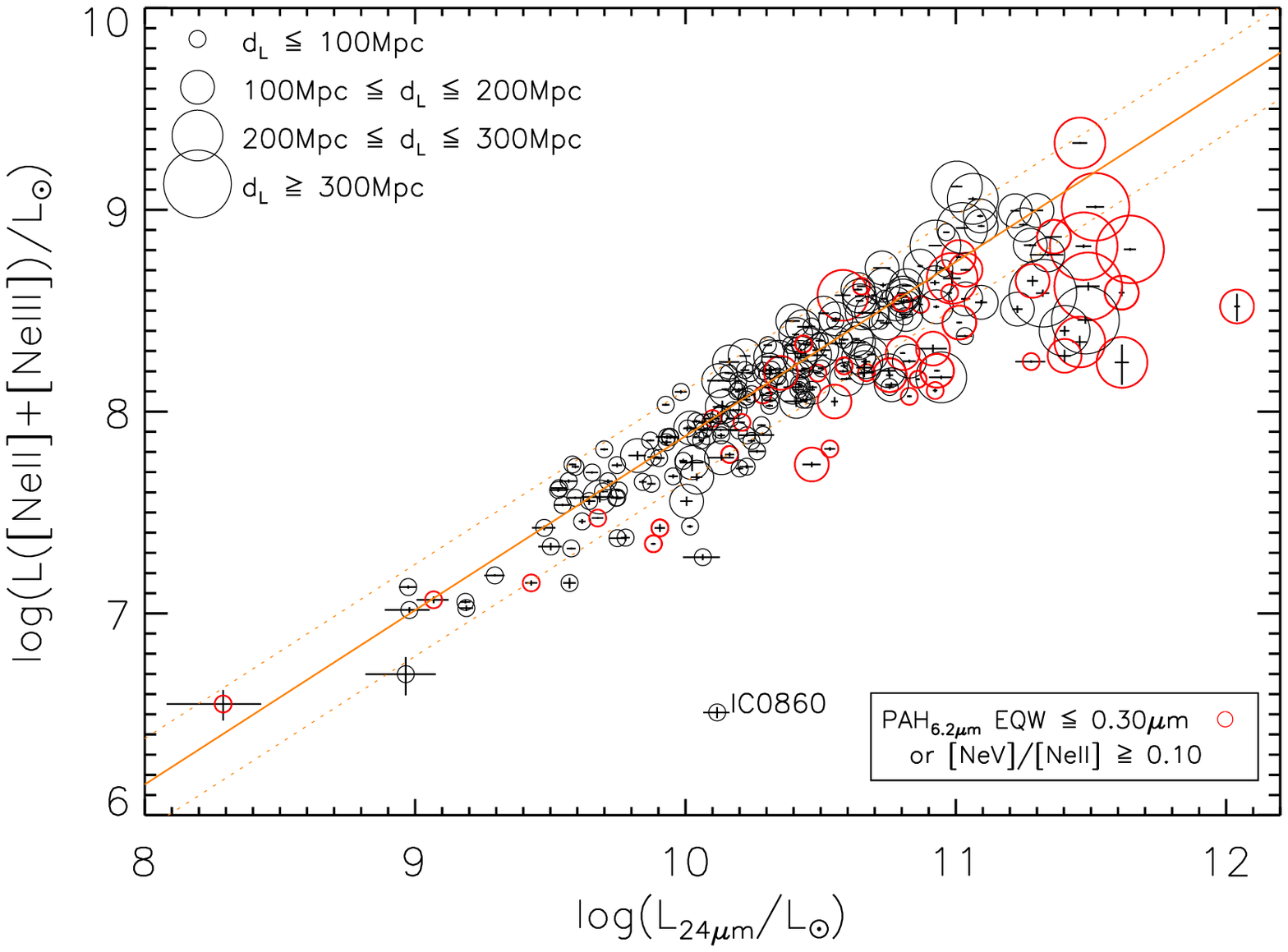}
    \caption { [Top] The total luminosity of the [NeII] and [NeIII]
      lines as a function of the infrared luminosity. Note that the
      luminosities of these neon lines of each source in a single
      system are summed for the comparison with the $L_{IR}$.  The AGN
      dominated sources (PAH$_{6.2 \mathrm{\mu m}}$ EQW $\leq 0.3 \,
      \mathrm{\mu m}$ or [NeV]/[NeII] $\geq 1.0$) are shown in the red
      circles. The luminosity distances of sources are denoted by symbol sizes
      (larger circle is more distant objects).  The orange solid line
      represents the best fit line when only the starburst dominated
      sources are considered. The best fit lines of the
      $L_{\mathrm{Ne}}$ vs. $L_{IR}$ relation estimated by \cite{Ho07}
      and \cite{Bern09} are also plotted in the gray dashed and dotted
      lines, respectively.  [Bottom] The luminosities of ${\rm [NeII]}
      + {\rm [NeIII]}$ vs. the synthetic $24\,\mathrm{\mu m}$
      luminosities.  The $24\,\mathrm{\mu m}$ luminosities are
      measured in the LH spectra.  The correlation line and its
      $1\sigma$ dispersion are represented by the orange solid and
      dotted lines, respectively.  }
    \label{fig:LIR_NeII_NeIII}
  \end{center}
\end{figure}

Since the [NeII] and [NeIII] lines typically have the strongest
emission in the $10-20\,{\rm \mu m}$ region, and their emitted fluxes
are correlated with $L_{IR}$ \citep{Ho07}, they can be used to estimate
star formation rates in starburst galaxies. 
The [NeII] and [NeIII] luminosity as a function of $L_{IR}$
\citep[Taken from Table 1 of][]{Armu09} for the GOALS sample is
presented in the top panel of Figure~\ref{fig:LIR_NeII_NeIII}.  In
this plot, for completeness, we include the sources whose mid-infrared
spectra appear to be dominated by AGN, and mark them in red.
The correlation between $L_{\rm [NeII]+[NeIII]}$ and $L_{IR}$ for the
GOALS sources is consistent with that seen by \cite{Ho07}, although
some objects are offset from the best fit line. These outliers are
mostly at the low- and high-luminosity end.  Excluding the AGN
dominated sources, the best fit line for the GOALS sample has a slope
of $0.95 \pm 0.06$ and an intercept of $-2.48 \pm 0.71$. Both slope
and intercept are consistent with the least-squares regression of
\cite{Ho07} over five orders of magnitude in the infrared
luminosity ($8 < \log{(L_{IR}/L_\odot)} < 13$).

In the bottom panel of Figure~\ref{fig:LIR_NeII_NeIII}, we show the
correlation of the ${\rm [NeII]} + {\rm [NeIII]}$ luminosity to the
$24\,\mu$m luminosity (calculated from the LH spectra using the MIPS
$24\,\mu$m response curve).  The $24\,\mu$m luminosity traces the warm
dust emission and can also be used to derive the star formation rate
in star-forming galaxies \citep{Calz07}.  Compared with $L_{IR}$, the
correlation between the neon line and $24\,\mu$m luminosity is
significantly tighter. The slope and the intercept of the starburst
dominated sources alone are $0.86 \pm 0.02$ and $-0.75 \pm 0.23$,
respectively. None of the sources at the low-luminosity end exhibits a
large scatter, except \object{IC~0860} (\object{IRAS~F13126+2453}),
which lies far below the regression line at $\log(L_{24\,{\rm \mu
    m}}/L_\odot) \sim 10.1$. At the high-luminosity end, the AGN
clearly fall below the correlation, most likely due to an excess of
warm dust emission, which is heated by the active nucleus.

A tighter correlation between $L_{{\rm [NeII]} + {\rm [NeIII]}}$ and
the synthetic $L_{24\,{\rm \mu m}}$, compared to $L_{{\rm [NeII]} +
  {\rm [NeIII]}}$ and $L_{IR}$, can be attributed to two factors.
First, warm dust emission detected in $L_{24\,{\rm \mu m}}$ is more
directly related to massive stars which heat the dust. Replacing the
$L_{IR}$ with $L_{24\,{\rm \mu m}}$ therefore lowers the scatter in
the correlation.  Second, the closest objects ($\leq 100$~Mpc), which
are mostly at the low-luminosity end, fall below the correlation line
in the upper panel of Figure~\ref{fig:LIR_NeII_NeIII}.  Varying
amounts of the total [NeII] and [NeIII] emission may therefore be
missed in the narrow IRS slit.  Using $L_{24\,{\rm \mu m}}$ provides a closer
match of the neon and warm dust emission, substantially improving the
correlation (the bottom panel in Figure~\ref{fig:LIR_NeII_NeIII}) for
the nearest LIRGs.

\begin{figure}
  \begin{center}
    \includegraphics[angle=0,scale=0.45]{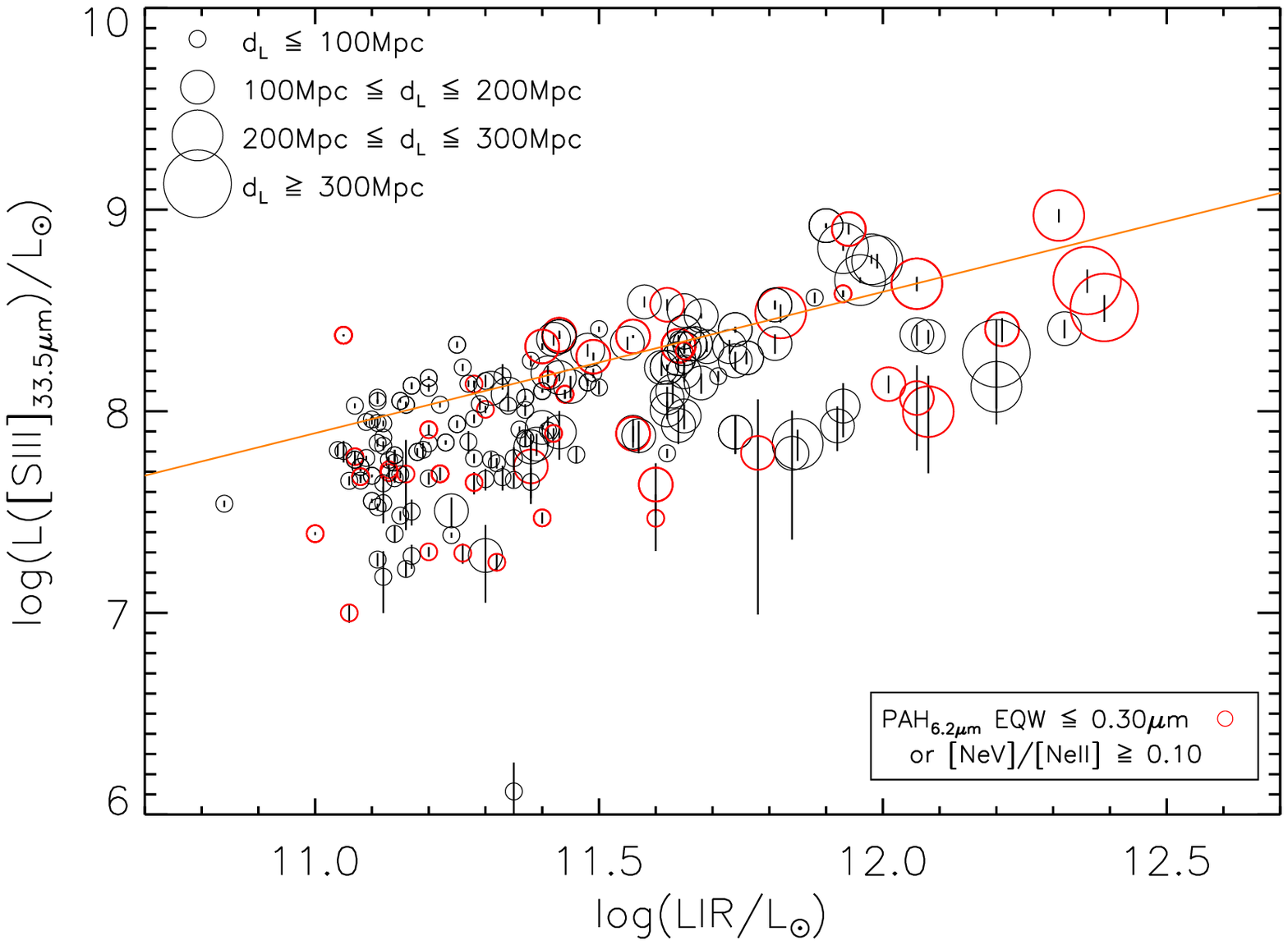}
    \includegraphics[angle=0,scale=0.45]{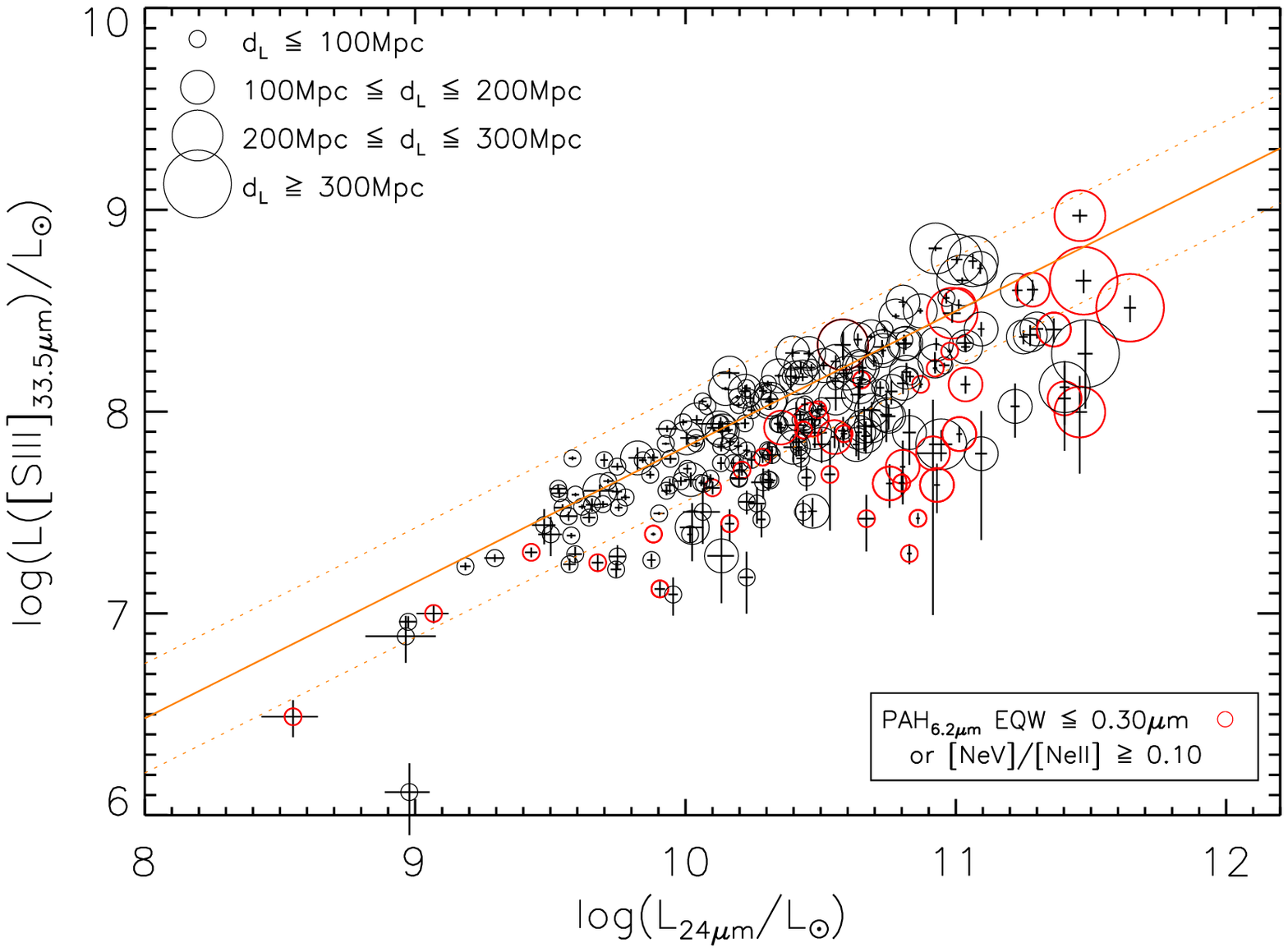}
    \caption { The same as in Figure~\ref{fig:LIR_NeII_NeIII} but for the
      [SIII] ($33.5\mu$m) luminosities.  For the comparison with the
      $L_{IR}$ (the top panel), the luminosities of the [SIII]
      emission line at $33.5\mu$m of each source in a single system
      (multiple nuclei) are summed.  }
    \label{fig:LIR_SIII}
  \end{center}
\end{figure}

In addition to the [NeII] and [NeIII] lines, the [SIII] $33.5{\rm \mu
  m}$ and [SiII] $34.8{\rm \mu m}$ lines can also be extremely strong
in the GOALS spectra.  Furthermore, since both lines are in the IRS LH
spectra, they can be directly compared to the $24\,\mu$m flux with no
potential aperture effects. The ionization potentials of $\rm S^{2+}$
and $\rm Si^{+}$ are $23.3 {\rm eV}$ and $8.2 {\rm eV}$, respectively.
Although the [SiII] emission line is generally stronger than
[SIII]$_{33.5{\rm \mu m}}$, the use of [SiII] for calibrating star
formation rates is problematic, because it may have a contribution
from older stars due to its low ionization potential.  Of course, this
is also true for the far-infrared emission in general, but much less
so in LIRGs and ULIRGs which tend to have much warmer emission
associated with the starburst.  In Figure~\ref{fig:LIR_SIII}, the
[SIII]$_{33.5{\rm \mu m}}$ luminosity is compared with $L_{IR}$ (upper
panel) and the synthetic $24\,\mu$m luminosities (lower panel).  The
AGN are again offset below the correlation for the starburst LIRGs.
As with the Ne lines, a tighter correlation is found for starbursts
than when $L_{24\,{\rm \mu m}}$ is used.  For the starburst dominated
sources alone, the best fit lines are $\log(L_{\rm [SIII]_{33.5 \mu
    m}}/L_\odot) = (0.70 \pm 0.08) \log(L_{IR}/L_\odot) + (0.18 \pm
0.93)$ and $\log(L_{\rm [SIII]_{33.5\mu m}}/L_\odot) = (0.67\pm
0.03)\log(L_{24\,\mathrm{\mu m}}/L_\odot) + (1.10\pm 0.27)$, for the
comparisons with the infrared luminosities and the synthetic
$24\,\mu$m luminosities, respectively.  The [SIII]$_{33.5{\rm \mu m}}$
emission line can also be used as a star formation rate indicator.
Although the correlation of $L_{\rm [SIII]_{33.5\mu m}}$
vs. $L_{24\,{\rm \mu m}}$ is not as tight as that of $L_{\rm
  [NeII]+[NeIII]}$ vs. $L_{24\,{\rm \mu m}}$, its scatter is as small
as that of the correlation in \cite{Ho07} between $L_{\rm
  [NeII]+[NeIII]}$ and $L_{IR}$ over five orders of magnitude in
$L_{IR}$.

\subsection{IR8 vs. Ionization and Linewidth}

Recent studies have revealed that the ratio of total infrared
luminosity ($L_{IR}$) to rest-frame $8\,{\rm \mu m}$ luminosity
($L_{8{\rm \mu m}}$), IR8, can be used to distinguish between ``main
sequence'' normal star-forming galaxies, and those undergoing a
``starburst'', characterized by high specific star formation rates
\citep{Dadd10b, Genz10, Elba11}.  High values of IR8 have also been
associated with compact starbursts \citep{Elba11}. With our
large sample of powerful starbursts in GOALS, we can examine possible
correlations of IR8 with other starburst properties.

\begin{figure}
  \begin{center}
    \includegraphics[angle=0,scale=0.45]{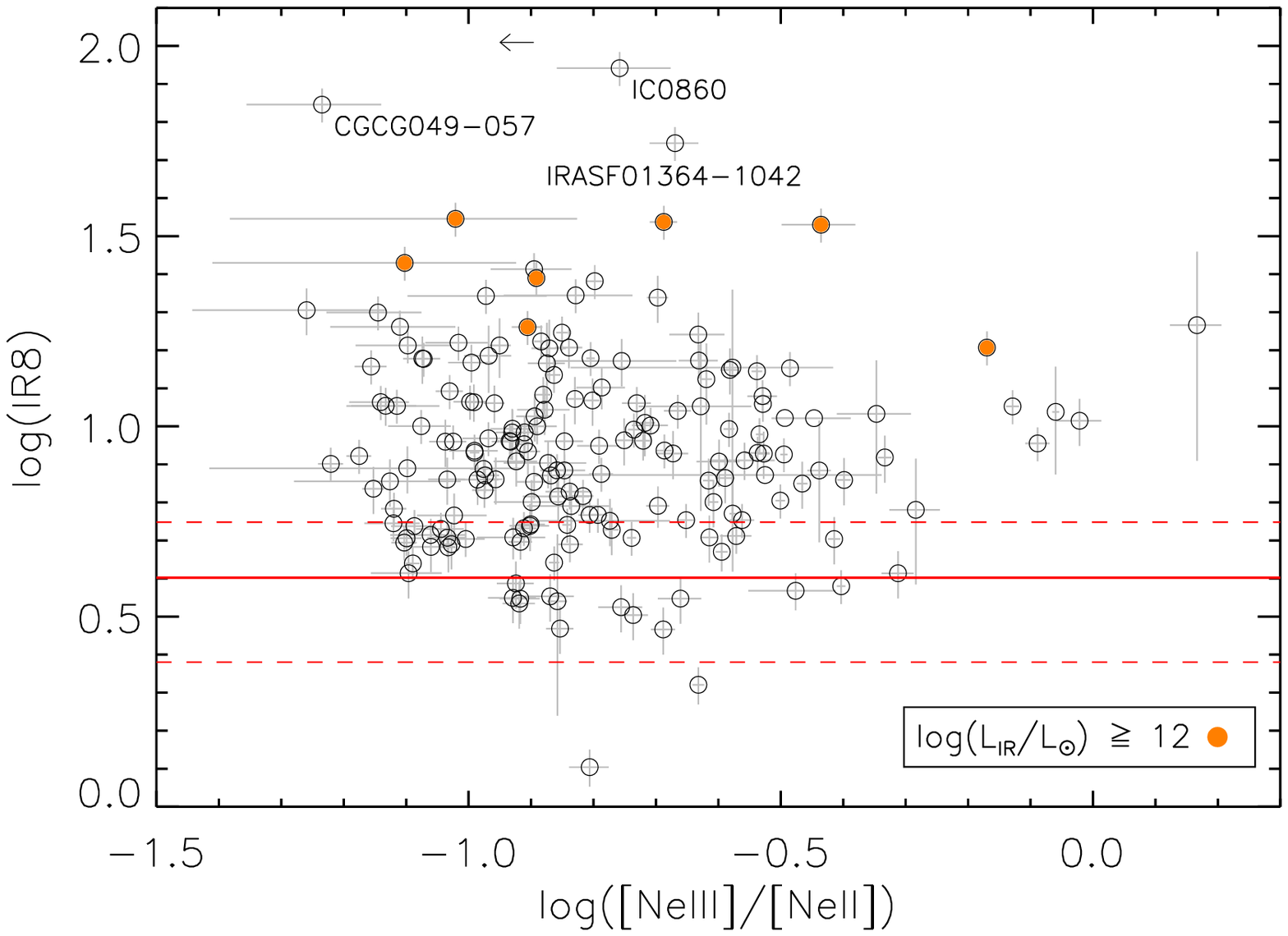}
    \includegraphics[angle=0,scale=0.45]{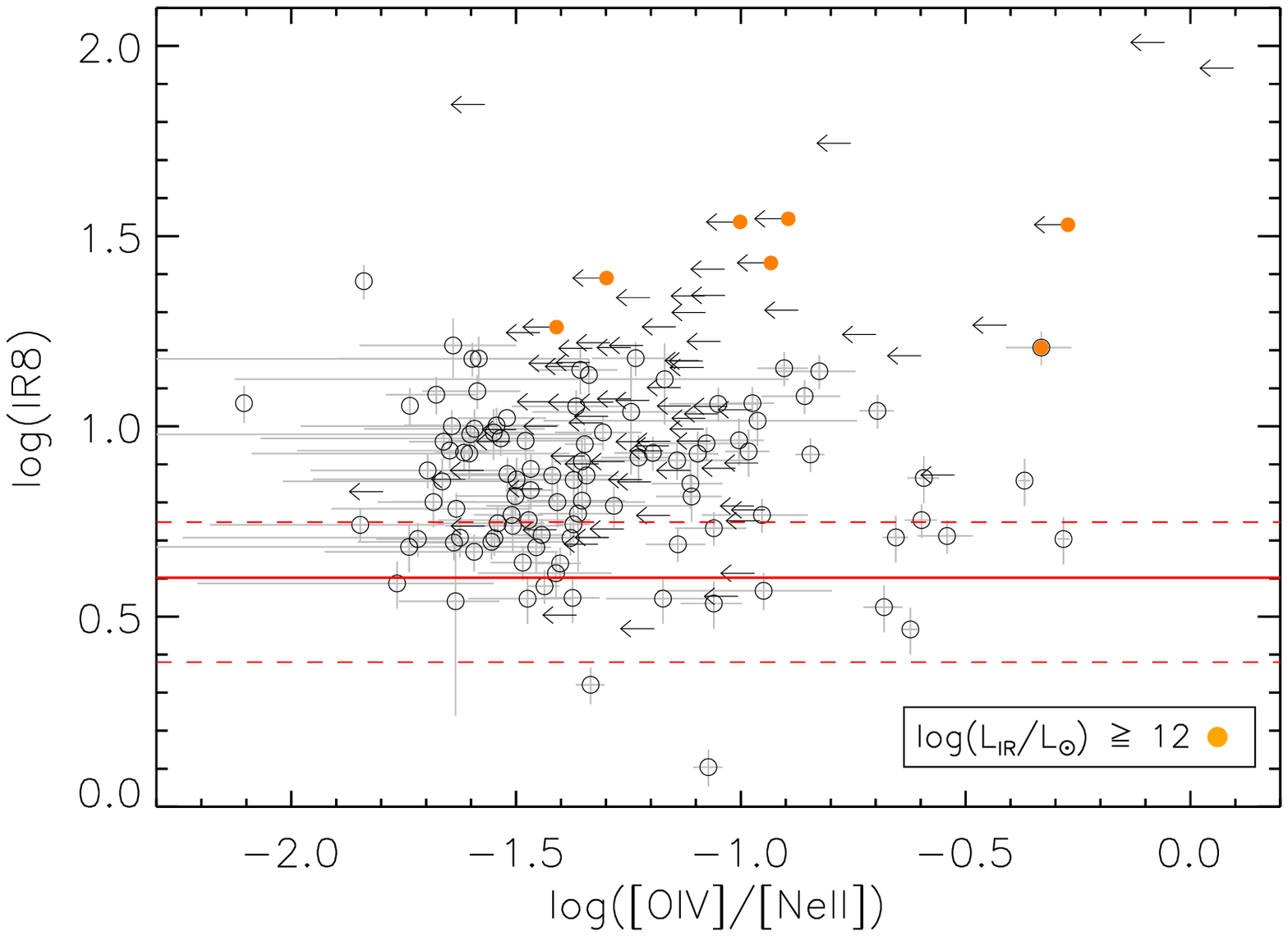}
    \caption{ IR8 ($L_{IR}/L_{8{\rm \mu m}}$) vs. [NeIII]/[NeII] (top)
      and [OIV]/[NeII] (bottom). Only starburst dominated LIRGs are
      shown in these figures. The left-facing arrows indicate upper
      limits of the line ratios. The orange filled symbols are
      ULIRGs. When a system has multiple nuclei, the $L_{IR}$ is
      divided proportionally to the fraction of their MIPS $70\,{\rm
        \mu m}$ emission. When the MIPS $70\,{\rm \mu m}$ image is not
      available, the $24\,{\rm \mu m}$ emission is used instead.  These
      figures do not include four nuclei in two systems, MCG+02-20-003
      and VV~250a, whose MIPS $24\,{\rm \mu m}$ images are not
      available. }
    \label{fig:IR8_line_ratio}
  \end{center}
\end{figure}

In Figure~\ref{fig:IR8_line_ratio}, we compare IR8 to the
[NeIII]/[NeII] and [OIV]/[NeII] ratios.  We note that all sources that
appear to be AGN dominated as classified at the beginning of
Seciton~\ref{sec:results} are excluded in these figures to minimize
AGN contamination in particular for the [OIV] line. The total infrared
luminosities are taken from \cite{Armu09} and the $8\,{\rm \mu m}$
luminosities are measured using the Spitzer IRAC $8\,{\rm \mu m}$
images.  Most GOALS sources lie well above the ``main sequence'' of
star-forming galaxies \citep[which has an average IR8 of
$3.9$,][]{Elba11}, consistent with them having compact, powerful
starbursts.  Although some of the sources with high [OIV]/[NeII]
limits also have large IR8 values, among the LIRGs with solid
detections, there is no correlation of either [NeIII]/[NeII] or
[OIV]/[NeII] with IR8.  Therefore, the LIRGs with the highest specific
star formation rates (sSFR) and the most compact starbursts do not
have a harder radiation field.  This lack of correlation may be due to
the different timescales probed by the IR8 ratio and high ionization
mid-infrared lines as traced by the [NeIII]/[NeII] or [OIV]/[NeII]
ratios.  The far-infrared flux, includes dust heated by stellar
populations with ages older than 10 Myr.  The [NeIII] and [OIV] lines,
however, trace the youngest stars, and as such they are not sensitive
to stellar populations older than about $5-6$~Myr. As a consequence,
changes in the line ratios can occur on much shorter timescales than
changes in IR8, if the starburst has been going for 10~Myr or
more. Alternatively, the lack of correlation may simply reflect the
fact that IR8 is more sensitive to the geometry of the region emitting
the (warm) dust, than the hardness of the radiation field as measured
in the HII regions themselves.

\begin{figure}
  \begin{center}
    \includegraphics[angle=0,scale=0.5]{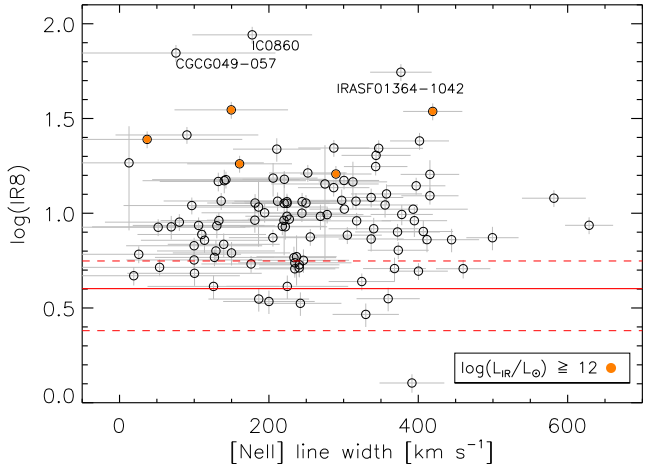}
    \caption{ The same as in Figure~\ref{fig:IR8_line_ratio} but plotted
      against to [NeII] line width (FWHM) corrected for the intrinsic
      velocity dispersion. }
    \label{fig:IR8_line_width}
  \end{center}
\end{figure}

Since we might expect the most compact, powerful starbursts to also be
the ones having the most turbulent ISM, we plot IR8 vs. the measured
line widths in Figure~\ref{fig:IR8_line_width}.  As with the
mid-infrared line flux ratios, we see no clear trend of line width and
IR8 among the GOALS sources.  The LIRGs with the broadest lines do not
appear farther from the star-forming main sequence of galaxies.  Even
though the IRS data are only sensitive to relatively high gas
velocities, the LIRGs with the largest IR8 seem to exhibit nearly the
full range in intrinsic line widths of the sample.  Therefore, the
compact starbursts seen in many LIRGs are not significantly increasing
the average ionized atomic gas motions -- at least not on the few
hundred ${\rm km\,s^{-1}}$ scales that can be probed with the high
resolution IRS spectra.

\section{Summary}\label{sec:summary}

We use the Starburst99 and the Mappings III models to diagnose the
physical properties and chemical abundances of the local LIRGs in the
GOALS sample.  Comparing mid-infrared diagrams of emission line ratios
with the photoionization starburst and shock-induced models, we find
that

\begin{enumerate}

\item Using the [SIV]/[NeII] and [NeIII]/[NeII] line flux ratios, we
  estimate starburst ages of $1 \lesssim \mathrm{Age\,[Myr]} \lesssim
  4.5$, and metallicities of $1 \lesssim Z\,[Z_\odot] \lesssim 2$,
  which are comparable with various optical metallicity measurements,
  showing that LIRGs are metal rich \citep[e.g.,][]{Rupk08}.  The
  electron densities of the GOALS starbusts range from one to a few
  hundred cm$^{-3}$, as measured from the [SIII] emission lines.  For
  the 91 starburst-dominated sources significantly above the low
  density limit of the [SIII] emission lines, we estimate mean and
  median electron densities of $340 \, \mathrm{cm^{-3}}$ and $300 \,
  \mathrm{cm^{-3}}$, respectively. Although the GOALS sources have a
  large dispersion in ionization parameter, most of the starburst
  sources are consistent with ionization parameters of $2\times10^7
  \lesssim q\,[\mathrm{cm\,s^{-1}}] \lesssim 8\times10^7$. Most of the
  LIRGs with $q > 8\times10^7 \, {\rm cm\,s^{-1}}$ appear to be AGN
  dominated (they have very small $6.2 {\rm \mu m}$ PAH EQW or strong
  [NeV] line emission relative to [NeII]).  However, there are five
  GOALS LIRGs which are very high ionization starbursts.  These
  sources could be slightly younger than the average LIRG.  We find no
  significant correlation of the fine structure line ratios with
  merger stage throughout the sample.

\item We have used the [SIV]/[NeII], [NeIII]/[NeII], [FeII]/[OIV], and
  [OIV]/[NeII] line flux ratios as well as their measured line profiles
  and line centroids to isolate LIRGs that might be experiencing large
  scale shocks.  We find that overall the number of shocked starburst
  LIRGs is small (10), and that the individual diagnostics often do
  not agree.  There appear to be five starburst LIRGs (including
  \object{NGC~6240}) that show evidence in at least one of these
  diagnostics for shock excitation and fast gas motions.  An
  additional five may have experienced shocks as indicated by their
  intense [FeII] emission, but they are now mostly ionized by young stars.

\item Resolved neon lines with FWHM $ > 600 \,\mathrm{km\,s^{-1}}$ (or
  $\gtrsim 330\,{\rm km\,s^{-1}}$ corrected for the intrinsic velocity
  dispersion) are seen in $\sim 30\%$ of the sample (80 objects).
  About $60\%$ of these (48 galaxies) are starburst dominated
  LIRGs. The [NeIII] linewidths of the AGN dominated sources are
  slightly greater, on average, than those of the starbursts.  At
  least 11 and as many as 19 LIRGs show a positive correlation of line
  width and ionization parameter, consistent with compact energy
  sources and possibly a stratified ISM.  Five GOALS sources show
  [NeIII] or [NeV] lines that are shifted by more than $200
  \,\mathrm{km\,s^{-1}}$ relative to [NeII], indicative of rapidly
  moving highly-ionized gas which may be closer to the ionizing source
  and participating in an outflow (or infall).

\item The [NeII]+[NeIII] luminosities of the starburst dominated
  sources show a correlation with $L_{IR}$. Although its dispersion is
  consistent with \cite{Ho07}, who found that this relation holds for
  over five magnitudes in infrared luminosity, the correlation becomes
  significantly tighter when the neon emission line luminosities are
  compared with the synthetic $24{\rm \mu m}$ luminosities measured in
  the LH spectra.  The luminosities of the ${\rm [SIII]_{33.5\mu m}}$
  emission line, which has a similar ionization potential to $\rm
  Ne^{+}$ and typically has a strong line flux like [NeII] and
  [NeIII], also correlate with $L_{IR}$ and $L_{24{\rm \mu m}}$. Thus
  $L_{\rm [SIII]_{33.5\mathrm{\mu m}}}$ can be used as a star
  formation rate indicator as well.

\item We do not find a correlation between $L_{IR}/L_{8\,{\rm \mu m}}$
  (IR8, the ratio of total infrared luminosity to rest-frame $8\,{\rm
    \mu m}$ luminosity) and the [NeIII]/[NeII] or [OIV]/[NeII]
  ratios. This implies that while IR8 may be tracing the compactness
  of dusty star formation \citep{Elba11} and the geometry of the warm
  dust emission in the starburst, it is not sensitive to the
  hardness of the ionizing radiation field, as probed by the
  mid-infrared atomic fine-structure emission line ratios. The more
  compact star-forming regions in starbursts are expected to also have
  the more turbulent ISM, which can be probed by line
  profiles. However, IR8 does not correlate with the [NeII] line width
  either.  The LIRGs with the largest IR8 seem to cover nearly
  the full range in line widths in the GOALS starburst sample.

\end{enumerate}


The mid-infrared fine-structure features detected with Spitzer IRS
high resolution spectroscopy, as presented here, provide an insight to
the basic physical properties of starburst and AGN dominated LIRGs in
the local Universe.  In conjunction with the forthcoming results of
[CII]$_{158\,{\rm \mu m}}$, [OI]$_{63\,{\rm \mu m}}$, and
[OIII]$_{88\,{\rm \mu m}}$ with the Herschel Space Observatory, these
results will lead to a more detailed understanding of the ionized and
neutral gas in local LIRGs.  Furthermore, the same techniques shown in
this paper can be applied using the higher resolution and more
sensitive instruments of the James Webb Space Telescope (JWST) and the
Space Infrared Telescope for Cosmology and Astrophysics (SPICA) in the
near future for understanding both low- and high-z galaxies.


\acknowledgments

The authors appreciate referee's useful suggestions which improved the
manuscript.  The authors are grateful to E. Levesque for providing the
stellar photo-ionization models. Hanae Inami thanks Grant-in-Aid for
Japan Society for the Promotion of Science (JSPS) Fellows (21-969) and
JSPS Excellent Young Researchers Overseas Visit Program for supporting
this work.  VC would like to acknowledge partial support from the EU
FP7 Grant PIRSES-GA-2012-316788.  The Spitzer Space Telescope is
operated by the Jet Propulsion Laboratory, California Institute of
Technology, under NASA contract 1407.  This research has made use of
the NASA/IPAC Extragalactic Database (NED) and the Infrared Science
Archive (IRSA) which are operated by the Jet Propulsion Laboratory,
California Institute of Technology, under contract with the National
Aeronautics and Space Administration.




{\it Facilities:} \facility{Spitzer}


\bibliography{IRSbib}

\begin{thebibliography}{78}
\expandafter\ifx\csname natexlab\endcsname\relax\def\natexlab#1{#1}\fi

\bibitem[{{Allen} {et~al.}(2008){Allen}, {Groves}, {Dopita}, {Sutherland}, \&
  {Kewley}}]{Alle08}
{Allen}, M.~G., {Groves}, B.~A., {Dopita}, M.~A., {Sutherland}, R.~S., \&
  {Kewley}, L.~J. 2008, \apjs, 178, 20

\bibitem[{{Armus} {et~al.}(2006){Armus}, {Bernard-Salas}, {Spoon}, {Marshall},
  {Charmandaris}, {Higdon}, {Desai}, {Hao}, {Teplitz}, {Devost}, {Brandl},
  {Soifer}, \& {Houck}}]{Armu06}
{Armus}, L., {et~al.} 2006, \apj, 640, 204

\bibitem[{{Armus} {et~al.}(2009){Armus}, {Mazzarella}, {Evans}, {Surace},
  {Sanders}, {Iwasawa}, {Frayer}, {Howell}, {Chan}, {Petric}, {Vavilkin},
  {Kim}, {Haan}, {Inami}, {Murphy}, {Appleton}, {Barnes}, {Bothun}, {Bridge},
  {Charmandaris}, {Jensen}, {Kewley}, {Lord}, {Madore}, {Marshall},
  {Melbourne}, {Rich}, {Satyapal}, {Schulz}, {Spoon}, {Sturm}, {U}, {Veilleux},
  \& {Xu}}]{Armu09}
---. 2009, \pasp, 121, 559

\bibitem[{{Bernard-Salas} {et~al.}(2009){Bernard-Salas}, {Spoon},
  {Charmandaris}, {Lebouteiller}, {Farrah}, {Devost}, {Brandl}, {Wu}, {Armus},
  {Hao}, {Sloan}, {Weedman}, \& {Houck}}]{Bern09}
{Bernard-Salas}, J., {et~al.} 2009, \apjs, 184, 230

\bibitem[{{Berta} {et~al.}(2011){Berta}, {Magnelli}, {Nordon}, {Lutz}, {Wuyts},
  {Altieri}, {Andreani}, {Aussel}, {Casta{\~n}eda}, {Cepa}, {Cimatti}, {Daddi},
  {Elbaz}, {F{\"o}rster Schreiber}, {Genzel}, {Le Floc'h}, {Maiolino},
  {P{\'e}rez-Fournon}, {Poglitsch}, {Popesso}, {Pozzi}, {Riguccini},
  {Rodighiero}, {Sanchez-Portal}, {Sturm}, {Tacconi}, \&
  {Valtchanov}}]{Berta11}
{Berta}, S., {et~al.} 2011, \aap, 532, A49

\bibitem[{{Binette} {et~al.}(1985){Binette}, {Dopita}, \& {Tuohy}}]{Bine85}
{Binette}, L., {Dopita}, M.~A., \& {Tuohy}, I.~R. 1985, \apj, 297, 476

\bibitem[{{Brandl} {et~al.}(2006){Brandl}, {Bernard-Salas}, {Spoon}, {Devost},
  {Sloan}, {Guilles}, {Wu}, {Houck}, {Weedman}, {Armus}, {Appleton}, {Soifer},
  {Charmandaris}, {Hao}, {Higdon}, \& {Herter}}]{Bran06}
{Brandl}, B.~R., {et~al.} 2006, \apj, 653, 1129

\bibitem[{{Calzetti} {et~al.}(2007){Calzetti}, {Kennicutt}, {Engelbracht},
  {Leitherer}, {Draine}, {Kewley}, {Moustakas}, {Sosey}, {Dale}, {Gordon},
  {Helou}, {Hollenbach}, {Armus}, {Bendo}, {Bot}, {Buckalew}, {Jarrett}, {Li},
  {Meyer}, {Murphy}, {Prescott}, {Regan}, {Rieke}, {Roussel}, {Sheth}, {Smith},
  {Thornley}, \& {Walter}}]{Calz07}
{Calzetti}, D., {et~al.} 2007, \apj, 666, 870

\bibitem[{{Chien} {et~al.}(2007){Chien}, {Barnes}, {Kewley}, \&
  {Chambers}}]{Chie07}
{Chien}, L.-H., {Barnes}, J.~E., {Kewley}, L.~J., \& {Chambers}, K.~C. 2007,
  \apjl, 660, L105

\bibitem[{{Crowther} {et~al.}(2002){Crowther}, {Dessart}, {Hillier}, {Abbott},
  \& {Fullerton}}]{Crow02}
{Crowther}, P.~A., {Dessart}, L., {Hillier}, D.~J., {Abbott}, J.~B., \&
  {Fullerton}, A.~W. 2002, \aap, 392, 653

\bibitem[{{Daddi} {et~al.}(2010){Daddi}, {Elbaz}, {Walter}, {Bournaud},
  {Salmi}, {Carilli}, {Dannerbauer}, {Dickinson}, {Monaco}, \&
  {Riechers}}]{Dadd10b}
{Daddi}, E., {et~al.} 2010, \apjl, 714, L118

\bibitem[{{Dale} {et~al.}(2006){Dale}, {Smith}, {Armus}, {Buckalew}, {Helou},
  {Kennicutt}, {Moustakas}, {Roussel}, {Sheth}, {Bendo}, {Calzetti}, {Draine},
  {Engelbracht}, {Gordon}, {Hollenbach}, {Jarrett}, {Kewley}, {Leitherer},
  {Li}, {Malhotra}, {Murphy}, \& {Walter}}]{Dale06}
{Dale}, D.~A., {et~al.} 2006, \apj, 646, 161

\bibitem[{{Dale} {et~al.}(2009){Dale}, {Smith}, {Schlawin}, {Armus},
  {Buckalew}, {Cohen}, {Helou}, {Jarrett}, {Johnson}, {Moustakas}, {Murphy},
  {Roussel}, {Sheth}, {Staudaher}, {Bot}, {Calzetti}, {Engelbracht}, {Gordon},
  {Hollenbach}, {Kennicutt}, \& {Malhotra}}]{Dale09}
---. 2009, \apj, 693, 1821

\bibitem[{{Dasyra} {et~al.}(2011){Dasyra}, {Ho}, {Netzer}, {Combes},
  {Trakhtenbrot}, {Sturm}, {Armus}, \& {Elbaz}}]{Dasy11}
{Dasyra}, K.~M., {Ho}, L.~C., {Netzer}, H., {Combes}, F., {Trakhtenbrot}, B.,
  {Sturm}, E., {Armus}, L., \& {Elbaz}, D. 2011, \apj, 740, 94

\bibitem[{{D{\'{\i}}az-Santos} {et~al.}(2007){D{\'{\i}}az-Santos},
  {Alonso-Herrero}, {Colina}, {Ryder}, \& {Knapen}}]{Diaz07}
{D{\'{\i}}az-Santos}, T., {Alonso-Herrero}, A., {Colina}, L., {Ryder}, S.~D.,
  \& {Knapen}, J.~H. 2007, \apj, 661, 149

\bibitem[{{D{\'{\i}}az-Santos} {et~al.}(2010){D{\'{\i}}az-Santos},
  {Charmandaris}, {Armus}, {Petric}, {Howell}, {Murphy}, {Mazzarella},
  {Veilleux}, {Bothun}, {Inami}, {Appleton}, {Evans}, {Haan}, {Marshall},
  {Sanders}, {Stierwalt}, \& {Surace}}]{Diaz10b}
{D{\'{\i}}az-Santos}, T., {et~al.} 2010, \apj, 723, 993

\bibitem[{{D{\'{\i}}az-Santos} {et~al.}(2011){D{\'{\i}}az-Santos},
  {Charmandaris}, {Armus}, {Stierwalt}, {Haan}, {Mazzarella}, {Howell},
  {Veilleux}, {Murphy}, {Petric}, {Appleton}, {Evans}, {Sanders}, \&
  {Surace}}]{Diaz11}
---. 2011, \apj, 741, 32

\bibitem[{{Dopita} {et~al.}(2000){Dopita}, {Kewley}, {Heisler}, \&
  {Sutherland}}]{Dopi00}
{Dopita}, M.~A., {Kewley}, L.~J., {Heisler}, C.~A., \& {Sutherland}, R.~S.
  2000, \apj, 542, 224

\bibitem[{{Draine}(2010)}]{Drai10}
{Draine}, B.~T. 2010, Physics of the Interstellar and Intergalactic Medium
  (Princeton University Press)

\bibitem[{{Elbaz} {et~al.}(2011){Elbaz}, {Dickinson}, {Hwang},
  {D{\'{\i}}az-Santos}, {Magdis}, {Magnelli}, {Le Borgne}, {Galliano},
  {Pannella}, {Chanial}, {Armus}, {Charmandaris}, {Daddi}, {Aussel}, {Popesso},
  {Kartaltepe}, {Altieri}, {Valtchanov}, {Coia}, {Dannerbauer}, {Dasyra},
  {Leiton}, {Mazzarella}, {Alexander}, {Buat}, {Burgarella}, {Chary}, {Gilli},
  {Ivison}, {Juneau}, {Le Floc'h}, {Lutz}, {Morrison}, {Mullaney}, {Murphy},
  {Pope}, {Scott}, {Brodwin}, {Calzetti}, {Cesarsky}, {Charlot}, {Dole},
  {Eisenhardt}, {Ferguson}, {F{\"o}rster Schreiber}, {Frayer}, {Giavalisco},
  {Huynh}, {Koekemoer}, {Papovich}, {Reddy}, {Surace}, {Teplitz}, {Yun}, \&
  {Wilson}}]{Elba11}
{Elbaz}, D., {et~al.} 2011, \aap, 533, A119

\bibitem[{{Evans} {et~al.}(2008){Evans}, {Vavilkin}, {Pizagno}, {Modica},
  {Mazzarella}, {Iwasawa}, {Howell}, {Surace}, {Armus}, {Petric}, {Spoon},
  {Barnes}, {Suer}, {Sanders}, {Chan}, \& {Lord}}]{Evan08}
{Evans}, A.~S., {et~al.} 2008, \apjl, 675, L69

\bibitem[{{Farrah} {et~al.}(2007){Farrah}, {Bernard-Salas}, {Spoon}, {Soifer},
  {Armus}, {Brandl}, {Charmandaris}, {Desai}, {Higdon}, {Devost}, \&
  {Houck}}]{Farr07}
{Farrah}, D., {et~al.} 2007, \apj, 667, 149

\bibitem[{{Genzel} {et~al.}(2010){Genzel}, {Tacconi}, {Gracia-Carpio},
  {Sternberg}, {Cooper}, {Shapiro}, {Bolatto}, {Bouch{\'e}}, {Bournaud},
  {Burkert}, {Combes}, {Comerford}, {Cox}, {Davis}, {Schreiber},
  {Garcia-Burillo}, {Lutz}, {Naab}, {Neri}, {Omont}, {Shapley}, \&
  {Weiner}}]{Genz10}
{Genzel}, R., {et~al.} 2010, \mnras, 407, 2091

\bibitem[{{Groves} {et~al.}(2008){Groves}, {Nefs}, \& {Brandl}}]{Grov08}
{Groves}, B., {Nefs}, B., \& {Brandl}, B. 2008, \mnras, 391, L113

\bibitem[{{Groves} {et~al.}(2004){Groves}, {Dopita}, \& {Sutherland}}]{Grov04a}
{Groves}, B.~A., {Dopita}, M.~A., \& {Sutherland}, R.~S. 2004, \apjs, 153, 9

\bibitem[{{Haan} {et~al.}(2011){Haan}, {Surace}, {Armus}, {Evans}, {Howell},
  {Mazzarella}, {Kim}, {Vavilkin}, {Inami}, {Sanders}, {Petric}, {Bridge},
  {Melbourne}, {Charmandaris}, {Diaz-Santos}, {Murphy}, {U}, {Stierwalt}, \&
  {Marshall}}]{Haan11a}
{Haan}, S., {et~al.} 2011, \aj, 141, 100

\bibitem[{{Hao} {et~al.}(2009){Hao}, {Wu}, {Charmandaris}, {Spoon},
  {Bernard-Salas}, {Devost}, {Lebouteiller}, \& {Houck}}]{Hao09}
{Hao}, L., {Wu}, Y., {Charmandaris}, V., {Spoon}, H.~W.~W., {Bernard-Salas},
  J., {Devost}, D., {Lebouteiller}, V., \& {Houck}, J.~R. 2009, \apj, 704, 1159

\bibitem[{{Hillier} \& {Miller}(1998)}]{Hill98}
{Hillier}, D.~J., \& {Miller}, D.~L. 1998, \apj, 496, 407

\bibitem[{{Ho} \& {Keto}(2007)}]{Ho07}
{Ho}, L.~C., \& {Keto}, E. 2007, \apj, 658, 314

\bibitem[{{Hopkins} {et~al.}(2008){Hopkins}, {Hernquist}, {Cox}, \& {Kere{\v
  s}}}]{HopkP08}
{Hopkins}, P.~F., {Hernquist}, L., {Cox}, T.~J., \& {Kere{\v s}}, D. 2008,
  \apjs, 175, 356

\bibitem[{{Houck} {et~al.}(2004){Houck}, {Roellig}, {van Cleve}, {Forrest},
  {Herter}, {Lawrence}, {Matthews}, {Reitsema}, {Soifer}, {Watson}, {Weedman},
  {Huisjen}, {Troeltzsch}, {Barry}, {Bernard-Salas}, {Blacken}, {Brandl},
  {Charmandaris}, {Devost}, {Gull}, {Hall}, {Henderson}, {Higdon}, {Pirger},
  {Schoenwald}, {Sloan}, {Uchida}, {Appleton}, {Armus}, {Burgdorf},
  {Fajardo-Acosta}, {Grillmair}, {Ingalls}, {Morris}, \& {Teplitz}}]{Houc04}
{Houck}, J.~R., {et~al.} 2004, \apjs, 154, 18

\bibitem[{{Howell} {et~al.}(2010){Howell}, {Armus}, {Mazzarella}, {Evans},
  {Surace}, {Sanders}, {Petric}, {Appleton}, {Bothun}, {Bridge}, {Chan},
  {Charmandaris}, {Frayer}, {Haan}, {Inami}, {Kim}, {Lord}, {Madore},
  {Melbourne}, {Schulz}, {U}, {Vavilkin}, {Veilleux}, \& {Xu}}]{Howe10}
{Howell}, J.~H., {et~al.} 2010, \apj, 715, 572

\bibitem[{{Inami} {et~al.}(2010){Inami}, {Armus}, {Surace}, {Mazzarella},
  {Evans}, {Sanders}, {Howell}, {Petric}, {Vavilkin}, {Iwasawa}, {Haan},
  {Murphy}, {Stierwalt}, {Appleton}, {Barnes}, {Bothun}, {Bridge}, {Chan},
  {Charmandaris}, {Frayer}, {Kewley}, {Kim}, {Lord}, {Madore}, {Marshall},
  {Matsuhara}, {Melbourne}, {Rich}, {Schulz}, {Spoon}, {Sturm}, {U},
  {Veilleux}, \& {Xu}}]{Inam10}
{Inami}, H., {et~al.} 2010, \aj, 140, 63

\bibitem[{{Iwasawa} {et~al.}(2011){Iwasawa}, {Sanders}, {Teng}, {U}, {Armus},
  {Evans}, {Howell}, {Komossa}, {Mazzarella}, {Petric}, {Surace}, {Vavilkin},
  {Veilleux}, \& {Trentham}}]{Iwas11}
{Iwasawa}, K., {et~al.} 2011, \aap, 529, A106

\bibitem[{{Kennicutt} {et~al.}(2003){Kennicutt}, {Armus}, {Bendo}, {Calzetti},
  {Dale}, {Draine}, {Engelbracht}, {Gordon}, {Grauer}, {Helou}, {Hollenbach},
  {Jarrett}, {Kewley}, {Leitherer}, {Li}, {Malhotra}, {Regan}, {Rieke},
  {Rieke}, {Roussel}, {Smith}, {Thornley}, \& {Walter}}]{Kenn03}
{Kennicutt}, Jr., R.~C., {et~al.} 2003, \pasp, 115, 928

\bibitem[{{Kewley} \& {Dopita}(2002)}]{Kewl02}
{Kewley}, L.~J., \& {Dopita}, M.~A. 2002, \apjs, 142, 35

\bibitem[{{Kewley} {et~al.}(2001){Kewley}, {Dopita}, {Sutherland}, {Heisler},
  \& {Trevena}}]{Kewl01}
{Kewley}, L.~J., {Dopita}, M.~A., {Sutherland}, R.~S., {Heisler}, C.~A., \&
  {Trevena}, J. 2001, \apj, 556, 121

\bibitem[{{Leitherer} {et~al.}(1999){Leitherer}, {Schaerer}, {Goldader},
  {Gonz{\'a}lez Delgado}, {Robert}, {Kune}, {de Mello}, {Devost}, \&
  {Heckman}}]{Leit99}
{Leitherer}, C., {et~al.} 1999, \apjs, 123, 3

\bibitem[{{Levesque} {et~al.}(2010){Levesque}, {Kewley}, \& {Larson}}]{Leve10}
{Levesque}, E.~M., {Kewley}, L.~J., \& {Larson}, K.~L. 2010, \aj, 139, 712

\bibitem[{{Lotz} {et~al.}(2008){Lotz}, {Jonsson}, {Cox}, \&
  {Primack}}]{Lotz08b}
{Lotz}, J.~M., {Jonsson}, P., {Cox}, T.~J., \& {Primack}, J.~R. 2008, \mnras,
  391, 1137

\bibitem[{{Lutz} {et~al.}(2003){Lutz}, {Sturm}, {Genzel}, {Spoon}, {Moorwood},
  {Netzer}, \& {Sternberg}}]{Lutz03}
{Lutz}, D., {Sturm}, E., {Genzel}, R., {Spoon}, H.~W.~W., {Moorwood}, A.~F.~M.,
  {Netzer}, H., \& {Sternberg}, A. 2003, \aap, 409, 867

\bibitem[{{Magnelli} {et~al.}(2013){Magnelli}, {Popesso}, {Berta}, {Pozzi},
  {Elbaz}, {Lutz}, {Dickinson}, {Altieri}, {Andreani}, {Aussel},
  {B{\'e}thermin}, {Bongiovanni}, {Cepa}, {Charmandaris}, {Chary}, {Cimatti},
  {Daddi}, {F{\"o}rster Schreiber}, {Genzel}, {Gruppioni}, {Harwit}, {Hwang},
  {Ivison}, {Magdis}, {Maiolino}, {Murphy}, {Nordon}, {Pannella}, {P{\'e}rez
  Garc{\'{\i}}a}, {Poglitsch}, {Rosario}, {Sanchez-Portal}, {Santini}, {Scott},
  {Sturm}, {Tacconi}, \& {Valtchanov}}]{Magn13}
{Magnelli}, B., {et~al.} 2013, \aap, 553, A132

\bibitem[{{Mihos} \& {Hernquist}(1994)}]{Miho94}
{Mihos}, J.~C., \& {Hernquist}, L. 1994, \apjl, 437, L47

\bibitem[{{Monreal-Ibero} {et~al.}(2010){Monreal-Ibero}, {Arribas}, {Colina},
  {Rodr{\'{\i}}guez-Zaur{\'{\i}}n}, {Alonso-Herrero}, \&
  {Garc{\'{\i}}a-Mar{\'{\i}}n}}]{Monr10}
{Monreal-Ibero}, A., {Arribas}, S., {Colina}, L.,
  {Rodr{\'{\i}}guez-Zaur{\'{\i}}n}, J., {Alonso-Herrero}, A., \&
  {Garc{\'{\i}}a-Mar{\'{\i}}n}, M. 2010, \aap, 517, A28

\bibitem[{{Moustakas} {et~al.}(2010){Moustakas}, {Kennicutt}, {Tremonti},
  {Dale}, {Smith}, \& {Calzetti}}]{Mous10}
{Moustakas}, J., {Kennicutt}, Jr., R.~C., {Tremonti}, C.~A., {Dale}, D.~A.,
  {Smith}, J., \& {Calzetti}, D. 2010, \apjs, 190, 233

\bibitem[{{Oliva} {et~al.}(1999{\natexlab{a}}){Oliva}, {Lutz}, {Drapatz}, \&
  {Moorwood}}]{Oliva99a}
{Oliva}, E., {Lutz}, D., {Drapatz}, S., \& {Moorwood}, A.~F.~M.
  1999{\natexlab{a}}, \aap, 341, L75

\bibitem[{{Oliva} {et~al.}(1999{\natexlab{b}}){Oliva}, {Moorwood}, {Drapatz},
  {Lutz}, \& {Sturm}}]{Oliva99b}
{Oliva}, E., {Moorwood}, A.~F.~M., {Drapatz}, S., {Lutz}, D., \& {Sturm}, E.
  1999{\natexlab{b}}, \aap, 343, 943

\bibitem[{{Pauldrach} {et~al.}(2001){Pauldrach}, {Hoffmann}, \&
  {Lennon}}]{Paul01}
{Pauldrach}, A.~W.~A., {Hoffmann}, T.~L., \& {Lennon}, M. 2001, \aap, 375, 161

\bibitem[{{Pereira-Santaella} {et~al.}(2010){Pereira-Santaella},
  {Diamond-Stanic}, {Alonso-Herrero}, \& {Rieke}}]{Pere10b}
{Pereira-Santaella}, M., {Diamond-Stanic}, A.~M., {Alonso-Herrero}, A., \&
  {Rieke}, G.~H. 2010, \apj, 725, 2270

\bibitem[{{Petric} {et~al.}(2011){Petric}, {Armus}, {Howell}, {Chan},
  {Mazzarella}, {Evans}, {Surace}, {Sanders}, {Appleton}, {Charmandaris},
  {D{\'{\i}}az-Santos}, {Frayer}, {Haan}, {Inami}, {Iwasawa}, {Kim}, {Madore},
  {Marshall}, {Spoon}, {Stierwalt}, {Sturm}, {U}, {Vavilkin}, \&
  {Veilleux}}]{Petr11}
{Petric}, A.~O., {et~al.} 2011, \apj, 730, 28

\bibitem[{{Pilyugin} \& {Thuan}(2005)}]{Pily05}
{Pilyugin}, L.~S., \& {Thuan}, T.~X. 2005, \apj, 631, 231

\bibitem[{{Rich} {et~al.}(2011){Rich}, {Kewley}, \& {Dopita}}]{Rich11}
{Rich}, J.~A., {Kewley}, L.~J., \& {Dopita}, M.~A. 2011, \apj, 734, 87

\bibitem[{{Rich} {et~al.}(2012){Rich}, {Torrey}, {Kewley}, {Dopita}, \&
  {Rupke}}]{Rich12}
{Rich}, J.~A., {Torrey}, P., {Kewley}, L.~J., {Dopita}, M.~A., \& {Rupke},
  D.~S.~N. 2012, \apj, 753, 5

\bibitem[{{Roche} \& {Aitken}(1984)}]{Roch84}
{Roche}, P.~F., \& {Aitken}, D.~K. 1984, \mnras, 208, 481

\bibitem[{{Rupke} {et~al.}(2010){Rupke}, {Kewley}, \& {Chien}}]{Rupk10b}
{Rupke}, D.~S.~N., {Kewley}, L.~J., \& {Chien}, L. 2010, \apj, 723, 1255

\bibitem[{{Rupke} {et~al.}(2008){Rupke}, {Veilleux}, \& {Baker}}]{Rupk08}
{Rupke}, D.~S.~N., {Veilleux}, S., \& {Baker}, A.~J. 2008, \apj, 674, 172

\bibitem[{{Salpeter}(1955)}]{Salp55}
{Salpeter}, E.~E. 1955, \apj, 121, 161

\bibitem[{{Sanders} {et~al.}(2003){Sanders}, {Mazzarella}, {Kim}, {Surace}, \&
  {Soifer}}]{Sand03}
{Sanders}, D.~B., {Mazzarella}, J.~M., {Kim}, D.-C., {Surace}, J.~A., \&
  {Soifer}, B.~T. 2003, \aj, 126, 1607

\bibitem[{{Savage} \& {Sembach}(1996)}]{Sava96}
{Savage}, B.~D., \& {Sembach}, K.~R. 1996, \araa, 34, 279

\bibitem[{{Seymour}(2009)}]{Seym09}
{Seymour}, N. 2009, in Panoramic Radio Astronomy: Wide-field 1-2 GHz Research
  on Galaxy Evolution

\bibitem[{{Smith} {et~al.}(2007){Smith}, {Draine}, {Dale}, {Moustakas},
  {Kennicutt}, {Helou}, {Armus}, {Roussel}, {Sheth}, {Bendo}, {Buckalew},
  {Calzetti}, {Engelbracht}, {Gordon}, {Hollenbach}, {Li}, {Malhotra},
  {Murphy}, \& {Walter}}]{Smit07}
{Smith}, J.~D.~T., {et~al.} 2007, \apj, 656, 770

\bibitem[{{Snijders} {et~al.}(2007){Snijders}, {Kewley}, \& {van der
  Werf}}]{Snij07}
{Snijders}, L., {Kewley}, L.~J., \& {van der Werf}, P.~P. 2007, \apj, 669, 269

\bibitem[{{Soto} \& {Martin}(2010)}]{Soto10}
{Soto}, K.~T., \& {Martin}, C.~L. 2010, \apj, 716, 332

\bibitem[{{Spoon} {et~al.}(2009){Spoon}, {Armus}, {Marshall}, {Bernard-Salas},
  {Farrah}, {Charmandaris}, \& {Kent}}]{Spoo09a}
{Spoon}, H.~W.~W., {Armus}, L., {Marshall}, J.~A., {Bernard-Salas}, J.,
  {Farrah}, D., {Charmandaris}, V., \& {Kent}, B.~R. 2009, \apj, 693, 1223

\bibitem[{{Spoon} \& {Holt}(2009)}]{Spoo09b}
{Spoon}, H.~W.~W., \& {Holt}, J. 2009, \apjl, 702, L42

\bibitem[{{Stierwalt} {et~al.}(2013){Stierwalt}, {Armus}, {Surace}, {Inami},
  {Petric}, {Diaz-Santos}, {Haan}, {Charmandaris}, {Howell}, {Kim}, {Marshall},
  {Mazzarella}, {Spoon}, {Veilleux}, {Evans}, {Sanders}, {Appleton}, {Bothun},
  {Bridge}, {Chan}, {Frayer}, {Iwasawa}, {Kewley}, {Lord}, {Madore},
  {Melbourne}, {Murphy}, {Rich}, {Schulz}, {Sturm}, {U}, {Vavilkin}, \&
  {Xu}}]{Stie13a}
{Stierwalt}, S., {et~al.} 2013, \apjs, 206, 1

\bibitem[{{Sturm} {et~al.}(2002){Sturm}, {Lutz}, {Verma}, {Netzer},
  {Sternberg}, {Moorwood}, {Oliva}, \& {Genzel}}]{Stur02}
{Sturm}, E., {Lutz}, D., {Verma}, A., {Netzer}, H., {Sternberg}, A.,
  {Moorwood}, A.~F.~M., {Oliva}, E., \& {Genzel}, R. 2002, \aap, 393, 821

\bibitem[{{Sturm} {et~al.}(2006){Sturm}, {Rupke}, {Contursi}, {Kim}, {Lutz},
  {Netzer}, {Veilleux}, {Genzel}, {Lehnert}, {Tacconi}, {Maoz}, {Mazzarella},
  {Lord}, {Sanders}, \& {Sternberg}}]{Stur06}
{Sturm}, E., {et~al.} 2006, \apjl, 653, L13

\bibitem[{{Sutherland} \& {Dopita}(1993)}]{Suth93}
{Sutherland}, R.~S., \& {Dopita}, M.~A. 1993, \apjs, 88, 253

\bibitem[{{Thornley} {et~al.}(2000){Thornley}, {Schreiber}, {Lutz}, {Genzel},
  {Spoon}, {Kunze}, \& {Sternberg}}]{Thor00}
{Thornley}, M.~D., {Schreiber}, N.~M.~F., {Lutz}, D., {Genzel}, R., {Spoon},
  H.~W.~W., {Kunze}, D., \& {Sternberg}, A. 2000, \apj, 539, 641

\bibitem[{{Tommasin} {et~al.}(2010){Tommasin}, {Spinoglio}, {Malkan}, \&
  {Fazio}}]{Tomm10}
{Tommasin}, S., {Spinoglio}, L., {Malkan}, M.~A., \& {Fazio}, G. 2010, \apj,
  709, 1257

\bibitem[{{Tremonti} {et~al.}(2004){Tremonti}, {Heckman}, {Kauffmann},
  {Brinchmann}, {Charlot}, {White}, {Seibert}, {Peng}, {Schlegel}, {Uomoto},
  {Fukugita}, \& {Brinkmann}}]{Trem04}
{Tremonti}, C.~A., {et~al.} 2004, \apj, 613, 898

\bibitem[{{V{\'a}zquez} \& {Leitherer}(2005)}]{Vazq05}
{V{\'a}zquez}, G.~A., \& {Leitherer}, C. 2005, \apj, 621, 695

\bibitem[{{Veilleux} {et~al.}(2002){Veilleux}, {Kim}, \& {Sanders}}]{Veil02}
{Veilleux}, S., {Kim}, D., \& {Sanders}, D.~B. 2002, \apjs, 143, 315

\bibitem[{{Veilleux} {et~al.}(2009){Veilleux}, {Rupke}, {Kim}, {Genzel},
  {Sturm}, {Lutz}, {Contursi}, {Schweitzer}, {Tacconi}, {Netzer}, {Sternberg},
  {Mihos}, {Baker}, {Mazzarella}, {Lord}, {Sanders}, {Stockton}, {Joseph}, \&
  {Barnes}}]{Veil09a}
{Veilleux}, S., {et~al.} 2009, \apjs, 182, 628

\bibitem[{{Verma} {et~al.}(2003){Verma}, {Lutz}, {Sturm}, {Sternberg},
  {Genzel}, \& {Vacca}}]{Verm03}
{Verma}, A., {Lutz}, D., {Sturm}, E., {Sternberg}, A., {Genzel}, R., \&
  {Vacca}, W. 2003, \aap, 403, 829

\bibitem[{{Whitmore} \& {Schweizer}(1995)}]{Whit95}
{Whitmore}, B.~C., \& {Schweizer}, F. 1995, \aj, 109, 960

\bibitem[{{Wu} {et~al.}(2006){Wu}, {Charmandaris}, {Hao}, {Brandl},
  {Bernard-Salas}, {Spoon}, \& {Houck}}]{Wu06}
{Wu}, Y., {Charmandaris}, V., {Hao}, L., {Brandl}, B.~R., {Bernard-Salas}, J.,
  {Spoon}, H.~W.~W., \& {Houck}, J.~R. 2006, \apj, 639, 157

\end{thebibliography}


\begin{deluxetable}{ccccccccc}
\tabletypesize{\scriptsize}
\tablewidth{0pc}
\tablecaption{SH Line Fluxes \label{tbl:spec_SH}}
\tablehead{
Object & RA (J2000) & Dec (J2000) & PA & [SIV] & [NeII] & [NeV] & [NeIII] & [SIII] \\
 & & &              & $10.5{\rm \mu m}$ & $12.8{\rm \mu m}$ & $14.3{\rm \mu m}$ & $15.6{\rm \mu m}$ & $18.7{\rm \mu m}$ 
}
\startdata
           NGC0023  &   00:09:53.280  &   25:55:26.698  &       -76.3197  &  $    0.88 \pm    0.74 $  &  $   62.17 \pm    0.93 $  &  $              < 0.83 $  &  $    9.48 \pm    0.34 $  &  $   23.94 \pm    0.54 $  \\   
           NGC0034  &   00:11:06.434  &  -12:06:26.003  &       -78.7347  &  $              < 1.95 $  &  $   52.83 \pm    1.18 $  &  $              < 1.59 $  &  $    6.99 \pm    0.65 $  &  $   10.47 \pm    1.24 $  \\   
            Arp256  &   00:18:50.759  &  -10:22:36.353  &       -79.5735  &  $    2.26 \pm    0.75 $  &  $   82.04 \pm    2.15 $  &  $              < 1.15 $  &  $   16.05 \pm    0.97 $  &  $   39.41 \pm    2.10 $  \\   
      ESO350-IG038  &   00:36:52.588  &  -33:33:17.757  &        119.642  &  $   44.82 \pm    0.35 $  &  $   33.47 \pm    0.41 $  &  $              < 1.15 $  &  $  103.60 \pm    0.63 $  &  $   45.68 \pm    1.12 $  \\   
        NGC0232\_W  &   00:42:45.666  &  -23:33:40.697  &       -88.0222  &  $              < 1.87 $  &  $   70.41 \pm    0.78 $  &  $    1.12 \pm    0.29 $  &  $    8.76 \pm    0.38 $  &  $   16.15 \pm    0.65 $  \\   
        NGC0232\_E  &   00:42:52.658  &  -23:32:27.487  &       -88.0338  &  $    3.74 \pm    0.39 $  &  $   12.55 \pm    0.34 $  &  $    4.00 \pm    0.25 $  &  $    9.07 \pm    0.35 $  &  $    6.01 \pm    0.37 $  \\   
\enddata
\tablecomments{ Spitzer/IRS Short-High line fluxes for the GOALS sources. The central wavelength of each line is indicated. The RA and DEC of the center of the slit is given for each source. Line fluxes are in units of $10^{-17}\,{\rm W\,m^{-2}}$. Sources without line fluxes or upper limits were either not observed with the Short-High module, or they have lines that fall outside the Short-High bandpass.  The other parameters used in this paper and their references are: $L_{IR}$ \citep[][Table 1]{Armu09}, $6.2 \mathrm{\mu m}$ PAH EQW and merger stages \citep[][Table 1]{Stie13a}, and IR8 (D\'iaz-Santos et al. in prep.). The full table is available in the electronic version of the paper.}
\end{deluxetable}
 
\begin{deluxetable}{cccccccccc}
\tabletypesize{\scriptsize}
\tablewidth{0pc}
\tablecaption{LH Line Fluxes \label{tbl:spec_LH}}
\tablehead{
Object & RA (J2000) & Dec (J2000) & PA & [SIII] & [OIV] & [FeII] & [SIII] & [SiII] & $L_{24\,{\rm \mu m}}$ \\
 & & & & $18.7{\rm \mu m}$ & $25.9{\rm \mu m}$   & $26.0{\rm \mu m}$  & $33.48{\rm \mu m}$  & $34.8{\rm \mu m}$ & 
}
\startdata
           NGC0023  &   00:09:53.322  &   25:55:22.955  &       -161.130  &  $   35.55 \pm    1.38 $  &  $    1.96 \pm    0.52 $  &  $    5.71 \pm    0.53 $  &  $   65.52 \pm    2.02 $  &  $  123.50 \pm    3.97 $  &  $    1.33 \pm    0.04 $  \\   
           NGC0034  &   00:11:06.484  &  -12:06:29.735  &       -163.545  &  $                 ... $  &  $              < 5.59 $  &  $              < 5.59 $  &  $             < 15.91 $  &  $   39.36 \pm    6.86 $  &  $    5.69 \pm    0.20 $  \\   
            Arp256  &   00:18:50.811  &  -10:22:40.084  &       -164.383  &  $   32.59 \pm    2.38 $  &  $    2.35 \pm    0.88 $  &  $    3.24 \pm    0.83 $  &  $   46.50 \pm    3.37 $  &  $   74.60 \pm    4.92 $  &  $    5.38 \pm    0.13 $  \\   
      ESO350-IG038  &   00:36:52.639  &  -33:33:14.027  &        34.8284  &  $   75.92 \pm    5.21 $  &  $              < 6.12 $  &  $    2.65 \pm    4.39 $  &  $   55.08 \pm    2.11 $  &  $   58.07 \pm    2.24 $  &  $    7.41 \pm    0.14 $  \\   
        NGC0232\_W  &   00:42:45.762  &  -23:33:44.269  &       -172.832  &  $   16.63 \pm    2.27 $  &  $    7.33 \pm    0.82 $  &  $    3.63 \pm    0.82 $  &  $   33.02 \pm    2.48 $  &  $   73.93 \pm    3.44 $  &  $    3.14 \pm    0.10 $  \\   
        NGC0232\_E  &   00:42:52.755  &  -23:32:31.060  &       -172.844  &  $    4.83 \pm    3.45 $  &  $   17.09 \pm    0.43 $  &  $              < 5.49 $  &  $    9.82 \pm    1.75 $  &  $   15.58 \pm    2.19 $  &  $    1.46 \pm    0.06 $  \\   
\enddata
\tablecomments{ Spitzer/IRS Long-High line fluxes for the GOALS sources. The central wavelength of each line is indicated. The RA and DEC of the center of the slit is given for each source. Line fluxes are in units of $10^{-17}\,{\rm W\,m^{-2}}$. The units of $L_{24\,{\rm \mu m}}$ is $10^{10} \, L_{\odot}$. Sources without line fluxes or upper limits were either not observed with the Long-High module, or they have lines that fall outside the Long-High bandpass. The other parameters used in this paper and their references are: $L_{IR}$ \citep[][Table 1]{Armu09}, $6.2 \mathrm{\mu m}$ PAH EQW and merger stages \citep[][Table 1]{Stie13a}, and IR8 (D\'iaz-Santos et al. in prep.). The full table is available in the electronic version of the paper.}
\end{deluxetable}
 

\clearpage

\appendix

\section{The Line Profiles of Neon and Oxygen Emission Lines}

\begin{figure}[h]
  \begin{center}
    \includegraphics[trim=0 0 0 0,angle=90,scale=0.6]{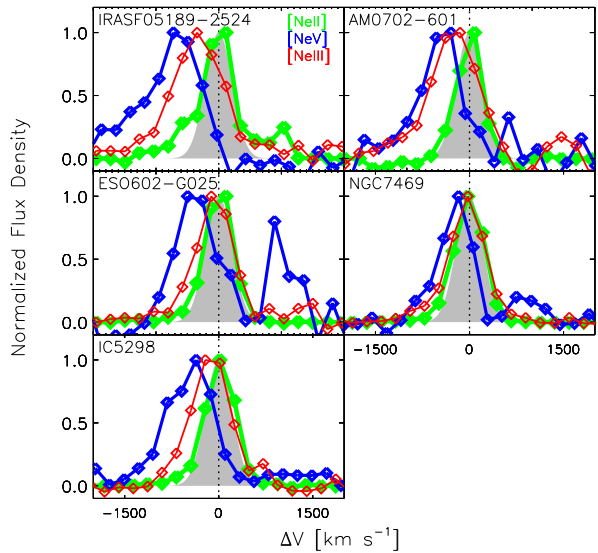}
    \caption { The [NeII] (green), [NeIII] (red), and [NeV] (blue)
      emission lines of which at least one of [NeIII] or [NeV] shows a
      velocity $\geq 200\,\mathrm{km\,s^{-1}}$ relative to the [NeII]
      velocity in the same SH spectra. The gray shades indicate the
      Gaussian profile of an resolved emission line at the IRS resolution of
      $R = 650 ~ (500\,{\rm km\,s^{-1}})$. }
    \label{fig:Ne_vel}
  \end{center}
\end{figure}

In Figure~\ref{fig:Ne_vel}, the five GOALS sources showing [NeIII] or [NeV] emission line
velocity shifts of $\geq 200\,{\rm km\,s^{-1}}$ relative to
the [NeII] emission are shown. All of these objects have asymmetric profiles
with blue wings.  Since all of these sources have [NeV]
detected, they are likely to harbor buried AGN.  However, not all of these sources are 
AGN dominated.  For example, \object{ESO~602-G025}
has ${\rm [NeV]/[NeII]} = 0.05 \pm 0.007$ and ${\rm EQW}(6.2\,{\rm \mu
  m}) = 0.45 \pm 0.01 \, {\rm \mu m}$, and thus most of the emission is probably 
  coming from a starburst.  Two of the sources
with high velocity shifts, \object{IRAS~F05189-2524}
($\log(L_{IR}/L_{\odot})=12.16$) and \object{NGC~7469}
($\log(L_{IR}/L_{\odot})=11.65$), are presented in
\cite{Spoo09b} and \cite{Dasy11}, respectively. \cite{Spoo09b}
show a clear correlation of blue shifted emission and ionization in
in local ULIRGs.  We find a similar result for the GOALS sources studied here.

\begin{figure}
  \begin{center}
    \includegraphics[angle=90,scale=0.6]{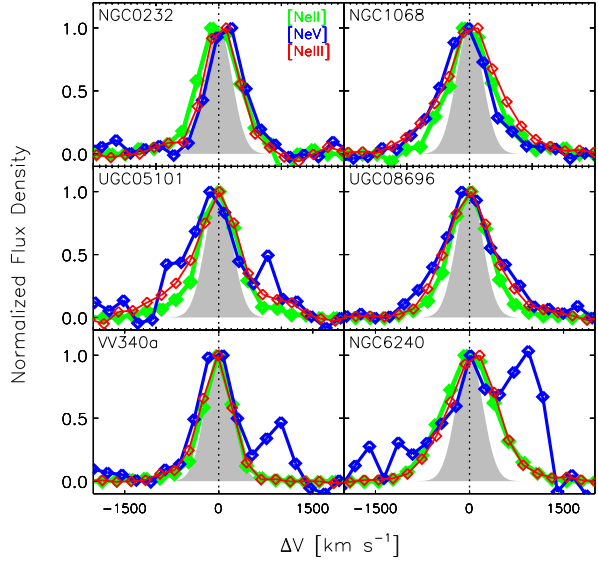}
    \caption { The same as Figure~\ref{fig:Ne_vel}, but now the
      sources with at least one of the neon lines showing a FWHM
      larger than $600\,\mathrm{km\,s^{-1}}$ are presented. In total
      there are 80 GOALS sources in this category. }
    \label{fig:Ne_width}
  \end{center}
\end{figure}

In Figure~\ref{fig:Ne_width}, representative objects possessing
resolved line profiles (FWHM $> 600\,{\rm km\,s^{-1}}$)~\footnote{This
  corresponds to a velocity dispersion of $330\,{\rm km\,s^{-1}}$
  corrected with the instrumental intrinsic velocity dispersion of
  $\sim 500\,{\rm km\,s^{-1}}$. This FWHM is $2\sigma$ over the
  average width of the unresolved neon emission line in the GOALS
  sources.} in at least one of the neon emission lines are
shown. There are 80 sources in total that have at least one resolved
neon emission line with FWHM $\geq 600\,{\rm km\,s^{-1}}$. All of the
sources show $\geq 200\,{\rm km\,s^{-1}}$ velocity shifts are in this
category as well.  Line widths, corrected for instrumental resolution,
are given in Table~\ref{tbl:width_SH}.

\begin{figure}
  \begin{center}
    \includegraphics[trim=0 0 0 0,angle=90,scale=0.6]{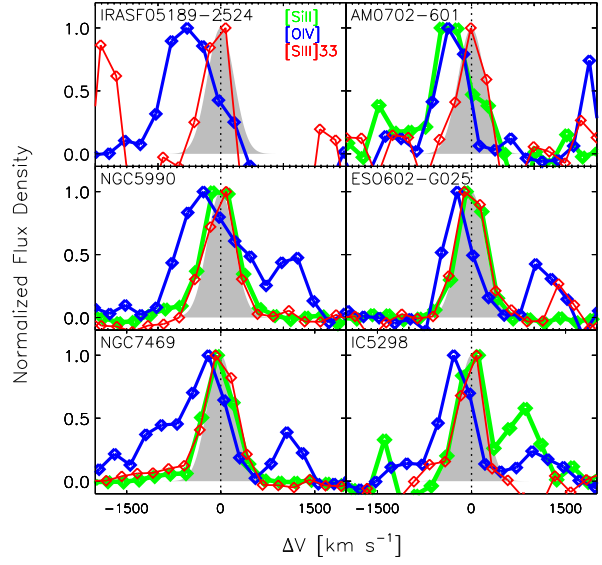}
    \caption { The same as Figure~\ref{fig:Ne_vel} but for the [SIII]
      at $33.5\mu$m (red), [SiII] (green), and [OIV] (blue) emission lines.}
    \label{fig:OIV_vel}
  \end{center}
\end{figure}

\begin{figure}
  \begin{center}
    \includegraphics[trim=0 0 0 0,angle=90,scale=0.6]{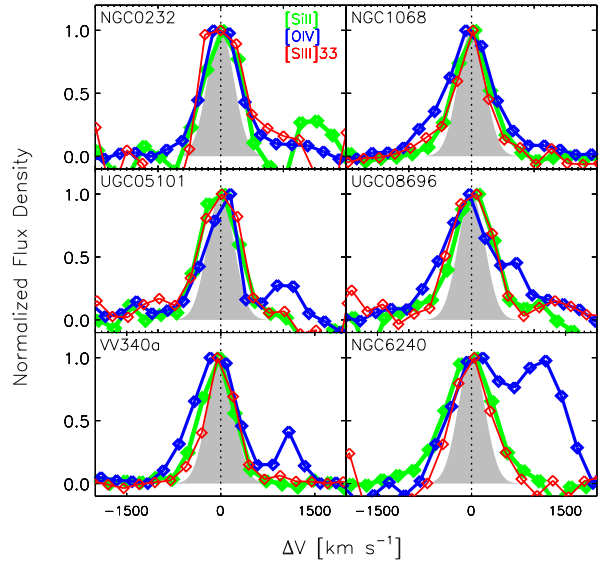}
    \caption { The same as Figure~\ref{fig:Ne_width} but for the
      [SIII] at $33.5\mu$m (red), [SiII] (green), and [OIV] (blue)
      emission lines.  The six representative examples out of total 98
      objects showing resolved [OIV] emission line (FWHM $\geq
      600\,\mathrm{km\,s^{-1}}$) are displayed in this figure.  }
    \label{fig:OIV_width}
  \end{center}
\end{figure}

As in Figure~\ref{fig:Ne_vel} and Figure~\ref{fig:Ne_width}, representative 
[OIV] emission line profiles are
shown in Figures~\ref{fig:OIV_vel} and \ref{fig:OIV_width}. 
There are ten objects in our sample exhibiting 
[OIV] emission lines shifted in velocity with respect to [NeII] by
more than $200 \, {\rm km \, s^{-1}}$.
There are 98 GOALS sources with resolved [OIV] emission lines, and 69 of these are 
starburst dominated.

\begin{deluxetable}{ccccccc}
\tabletypesize{\small}
\tablewidth{0pc}
\tablecaption{Neon Emission Line Widths for the Sources Showing Resolved Lines \label{tbl:width_SH}}
\tablehead{
Object & [NeII] & [NeIII] & [NeV]
}
\startdata
           NGC0034  &  $     355 \pm      26 $  &  $     342 \pm      45 $  &                       ... \\   
        NGC0232\_W  &                       ... &  $     345 \pm      30 $  &                       ... \\   
        NGC0232\_E  &  $     554 \pm      31 $  &  $     436 \pm      30 $  &  $     492 \pm      55 $  \\   
          NGC0317B  &                       ... &  $     338 \pm      28 $  &                       ... \\   
     MCG-03-04-014  &                       ... &  $     394 \pm      34 $  &                       ... \\   
          RR032\_N  &                       ... &  $     401 \pm      65 $  &                       ... \\   
\enddata
\tablecomments{Resolved neon emission line widths for the 80 sources which show at least one resolved neon emission line features. More can be found in the electronic version of the paper. Intrinsic linewidths, corrected for instrumental broadening, are given in ${\rm km\,s^{-1}}$. See text for a full description.}
\end{deluxetable}
 

\end{document}